\documentclass[
  journal=medium,
  manuscript=article-type,
  year=2026,
  volume=1,
]
{cup-article}

\def\og{``}
\def\fg{''}
\usepackage{soul}
\usepackage[T1]{fontenc}
\usepackage{lmodern}

\usepackage{amsmath, amssymb, amsthm}
\usepackage{mathtools}
\usepackage[nopatch]{microtype}
\usepackage{booktabs}
\usepackage{graphicx}
\graphicspath{{figures/}}
\usepackage{xcolor}
\usepackage{subcaption}
\usepackage[hidelinks]{hyperref}
\usepackage{longtable} 

\makeatletter
\def\cup@journal@name{Version 1.6 --- 2026-11-12}
\makeatother

\newtheorem{proposition}{Proposition}

\newtheorem{definition}{Definition}
\newtheorem{remark}{Remark}

\newcommand{\R}{\mathbb{R}}
\newcommand{\E}{\mathbb{E}}

\newcommand{\St}{\mathrm{St}}
\newcommand{\sym}{\mathrm{sym}}
\newcommand{\norm}[1]{\left\|#1\right\|}

\def\calF{\mathcal{F}}

\def\Esp{\mathbb{E}}

\def\narr#1{{\sc #1}}

\title{The Algebra of Narratives and the Monitoring of Their Relative Dominance}

\author{Juliette Anglade}
\affiliation{École polytechnique, Institut Polytechnique de Paris, Palaiseau, France}

\author{Charles-Albert Lehalle}
\affiliation{CMAP, CNRS, École polytechnique, Institut Polytechnique de Paris, Palaiseau, France}
\email[Charles-Albert Lehalle]{charles-albert.lehalle@polytechnique.edu}

\author{Mario Massy}
\affiliation{École polytechnique, Institut Polytechnique de Paris, Palaiseau, France}


\begin{document}

\begin{abstract}
We propose an algebraic methodology for measuring the relative prevalence of competing
narratives within a time-evolving corpus of texts.
Standard cosine similarity between article embedding and narrative vectors fails to
isolate the idiosyncratic contribution of each narrative, because narratives sharing a
common topic produce strongly correlated vector representations.
We address this by orthogonalizing the narrative vectors via a global max-min
optimization on the Stiefel manifold, which neutralizes shared semantic components
while preserving each narrative's specific direction.
We ground this geometric procedure in a linear generative model for article embedding
that provides an unbiased estimator of narrative prevalence and formally explains why
the naive approach suffers from a correlation bias.
Empirical validation on the publicly available News Category Dataset
(HuffPost, 2012--2018) confirms that the orthogonalized estimator recovers
narrative-specific dynamics aligned with known historical events, whereas
the naive cosine projection yields synchronized, uninformative time series.
\end{abstract}

\section{Positioning}
\label{sec:introbis}

Since the emergence of vector representations of texts, their power has been evidenced
by the structure of distance in the Euclidean space to which they belong.
Early demonstrations include the native clustering of Self-Organizing Map (SOM)
prototypes on Grimm tales~\autocite{honkela1995contextual} and the celebrated
\textit{king} $-$ \textit{man} $+$ \textit{woman} $=$ \textit{queen} relation
satisfied by word2vec embedding~\autocite{mikolov2013linguistic}.
Modern bidirectional encoders such as BERT have turned semantic similarity into
a practical primitive, notably in dense information retrieval~\autocite{wang2024utilizing}
and Retrieval-Augmented Generation (RAG) pipelines~\autocite{lewis2020retrieval}.
Concurrently, the concept of \emph{narrative competition} has gained traction in
economic literature~\autocite{shiller2020narrative}:
narratives are the mean field through which opinions form,
and agents condition their decisions on the perceived relative dominance of narratives.
Institutional actors exploit this dynamic; for instance, central banks
actively shape communication narratives to anchor public and market
expectations~\autocite{haldane2017conversation,macaulay2023narrative,hansen2018transparency}.
Market participants continuously probe and reshape narratives in a closed feedback loop
that renders straightforward causal analysis difficult.

The narrative viewpoint differs fundamentally from sentiment analysis.
Rather than classifying a document along a coarse positive--negative polarity spectrum---an approach that
conflates word frequency with topical relevance~\autocite{pang2008opinion,loughran2011liability}---it
uses the alignment of a text with the vector representation of a narrative as
a measure of \emph{importance}.
Formally, the cosine similarity between a news article and a narrative takes values
close to $+1$ when they are perfectly aligned, close to $-1$ when contradictory,
and close to zero when semantically independent in the high-dimensional embedding space,
although uncalibrated cosine similarities in dense representations are subject to known geometric distortions~\autocite{steck2024cosine}.

This paper addresses the following question: how can one monitor, over time and
at scale, the relative prevalence of competing narratives within a large corpus of
text? Existing approaches largely fall into three families (see~\autocite{gentzkow2019text} for a survey of text methods in economics). The first relies on keyword or dictionary-based counting: canonical
examples include the Economic Policy Uncertainty index of~\autocite{baker2016measuring},
built from keyword counts in newspaper archives, and the finance-specific sentiment
dictionaries of~\autocite{loughran2011liability}. Being built on exact vocabulary
matches, this family offers limited semantic granularity: a narrative expressed
through paraphrase or implication, without the target keywords, goes undetected.
The second relies on unsupervised topic modeling, typically via Latent Dirichlet
Allocation applied to financial or economic news~\autocite{bybee2020structure,larsen2019business},
or, closer to our setting, on text-semantics pipelines that extract narrative
statements from syntactic structure~\autocite{ash2024relatio}. Although richer than
raw keyword counts, this family remains fundamentally a bag-of-words approach: topics
are inferred from word co-occurrence statistics across the corpus, with no representation
of synonymy or paraphrase and no contextualization of polysemous words, and more
importantly for our purposes, a narrative of interest cannot be specified \emph{a
priori} by a short description; it must instead be statistically prevalent enough in
the corpus to emerge as a topic on its own. The third relies on supervised classifiers
trained to detect a predefined narrative from labeled examples: for
instance,~\autocite{coan2021computer} build a two-level taxonomy of climate-change
contrarian claims and train a classifier on tens of thousands of manually annotated
paragraphs.
Unlike the first two families, such classifiers, built on contextual
embedding, can achieve fine-grained semantic resolution but only at the cost of
substantial human labeling effort for every narrative of interest, and they do not
generalize to a newly introduced narrative without retraining.

The qualitative leap we seek is the ability to measure narrative prevalence at a
fine level of semantic nuance without either family's cost: neither reduced to a
fixed vocabulary or a corpus-driven topic, nor requiring labeled examples per
narrative. Just as importantly, we seek to compare the resulting measurements
\emph{across} narratives. This
requires prevalence estimates for different narratives to sit on a common,
mutually comparable scale, not merely to be individually well-behaved in time.
The methodology we propose specifies a narrative \emph{a priori} by a short
descriptive text, and measures its prevalence directly against an unlabeled
corpus, in the spirit of zero-shot, embedding-based text
classification~\autocite{schopf2023evaluating} and complementary to LLM-guided narrative extraction~\autocite{schmidt2025identifying}.

Measuring relative narrative strength poses the following challenge. Even apparently opposed narratives share a common semantic direction that causes
    their cosine scores to rise and fall in unison for reasons unrelated to their proper meaning, reflecting the pervasive anisotropy of dense language models~\autocite{ethayarajh2019contextual,mu2018all}.
    Disentangling \emph{idiosyncratic} from \emph{common} variation is therefore essential.

\paragraph{Contributions.}
We propose an algebraic framework that addresses this challenge. Our key contributions are:
\begin{itemize}
    \item We define the concept of \textit{pure narratives}, a way to model narratives that retain only their idiosyncratic meaning, separate from whatever is common in the collection.These are represented as a family of orthonormal vectors in the embedding space. We then propose a linear generative model for text embedding (typically when the text is an article), that provides a statistical foundation for demonstrating the need for pure narratives (rather than raw embedding obtained by encoding texts through a language model) to measure narrative prevalence and leads to an unbiased estimator of prevalence. (Section ~\ref{sec:model})
    \item Building on this diagnosis, we introduce a global orthogonalization procedure on the
          Stiefel manifold that constructs, from an arbitrary collection of
          candidate narrative texts, the corresponding collection of pure
          narratives: for each candidate, a representative vector that is close
          to the original text while being orthogonal to all others
          (Section~\ref{sec:ortho}).
    \item We perform an empirical validation on a publicly available labeled news dataset,
          including a counterfactual study contrasting the naive and
          orthogonalized measures, demonstrating qualitative and quantitative
          agreement with known historical events (Section~\ref{sec:application}).
\end{itemize}

It is not the first time algebra has been used in the space of embedding to remove biases or common effects. This has been notably done to remove gender biases from language models~\autocite{bolukbasi2016man,dev2019attenuating,dev2021oscar,dawkins2024projective}. The main difference is that whereas such operations on gender bias has to pre-defined polarities to orthogonalize too (the vector of the ``he–she difference'' or a set of a gender-paired sentences), in the case of a collation of narratives the directions to orthogonalize to are only implicitly defined. 
That is why we have to turn to the Stiefel manifold: looking for representative of narratives that are as close as the original as possible but as orthogonal to the rest of others as possible.

\section{A Generative Model for Text Embedding when Narratives are Involved}
\label{sec:model}

\subsection{Pure Narratives: an Idealized Abstraction}
\label{sec:narratives:embedding}

We take a \emph{narrative} to be an abstract idea, claim, or framing about the
world: an assertion, an interpretation of events, or a story that can be told
about them. As a result, a piece of text can be a \emph{representation} of a
given narrative without being the narrative itself. Text is not the only
possible representation, however. Language models make it possible to build a
bridge between the qualitative world of sentences and words and the
quantitative world. A narrative can then be represented as
a vector in a latent space $\mathcal H \subset \mathbb R^d$, so that comparing
or measuring narratives reduces to standard geometric operations on
$\mathcal H$.  This is a truly Platonist
claim: just as Plato's Ideas live in their own world and language is only a
support for them, \emph{narratives} live in their own world, and texts or
vectors are only supports for them.

However, by abuse of terminology, we will sometimes identify a narrative
with its representation (either a text or a vector). Throughout the paper, we work with two structured collections of twelve narratives (four thematic chapters of three narratives each):
\begin{itemize}
    \item a \textbf{Contemporary Policy Collection} (Appendix~\ref{app:narratives_illustration}, Table~\ref{tab:narratives}) spanning four modern macro-themes (\emph{Climate}, \emph{Artificial Intelligence}, \emph{Geopolitics}, and \emph{Economics}), which we use in Sections~\ref{sec:model}--\ref{sec:ortho} to illustrate embedding anisotropy and the geometry of Stiefel versus Gram--Schmidt orthogonalization; and
    \item a \textbf{Historical HuffPost Validation Collection} (Appendix~\ref{app:narratives}, Table~\ref{tab:narratives2}) spanning the four main labeled categories of the 2012--2018 \emph{News Category Dataset} (\emph{Politics}, \emph{Entertainment}, \emph{Parenting}, and \emph{Travel}), which we use in Section~\ref{sec:application} for historical time-series and counterfactual category validation.
\end{itemize}

For instance, using the first narrative of our Contemporary Policy Collection (Appendix~\ref{app:narratives_illustration}), the left panel of Table~\ref{tab:cosine_similarity} shows the cosine similarities (for \emph{ModernBERT} embeddings) of three English formulations of \narr{Extreme Heat and Emergency Policy Response}:
\begin{enumerate}
    \item \narr{\small Record-breaking temperatures across southern Europe this summer triggered unprecedented heatwaves, forcing governments to activate emergency climate plans and accelerate the transition toward renewable energy sources in order to meaningfully reduce carbon emissions and limit further global warming.}
    \item \narr{\small Record-breaking temperatures across southern Europe this summer fueled unprecedented heatwaves, prompting governments to implement emergency climate measures and fast-track the adoption of renewable energy to curb carbon emissions and mitigate further global warming.}
    \item \narr{\small Southern Europe faced extreme heat this summer, with record-breaking temperatures triggering a climate emergency and driving governments to intensify policy responses, accelerate the shift to clean energy, and implement measures to reduce carbon emissions and limit future global warming.}
\end{enumerate}
Their pairwise similarities are high ($0.95$--$0.99$), which explains why embedding vectors, despite living in a high-dimensional space ($d = 768$ for \emph{ModernBERT-base}), can serve as algebraic representatives of the narrative itself.

Nevertheless, if we take two other narratives from distinct chapters of the same collection (Appendix~\ref{app:narratives_illustration}):
\begin{itemize}
    \item[4.] \narr{NATO Eastern Flank and the Russia--Ukraine War} (\emph{Geopolitics}): \narr{\small The ongoing military conflict between Russia and Ukraine continues to destabilize the European geopolitical order, compelling NATO allies to reinforce their eastern flank with additional troops and to significantly increase national defense budgets in response to sustained and escalating Russian aggression.}
    \item[5.] \narr{IMF Forecasts and Global Growth Slowdown} (\emph{Economics}): \narr{\small Global economic growth is decelerating at a concerning pace according to the latest International Monetary Fund projections, which identify trade tensions, rising public debt levels, and slowing emerging market performance as the primary drivers behind deteriorating worldwide economic prospects.}
\end{itemize}
the right panel of Table~\ref{tab:cosine_similarity} shows that their pairwise cosine similarity is only slightly lower (around $0.91$--$0.93$ instead of $0.95$--$0.99$), remaining above $0.90$. It is due to common English grammar, sentence structure, and embedding anisotropy~\autocite{ethayarajh2019contextual,mu2018all}. 
Structural semantic similarities mask the underlying differences between narratives.

This observation of apparent similarity justifies the following definition.

\begin{table}
\caption{Pairwise Cosine Similarity Matrix (ModernBERT) for the narratives of the example inspired from Appendix~\ref{app:narratives_illustration}.}
\label{tab:cosine_similarity}
\centering
\begin{tabular}{lrrr c| lrrr}
\toprule
 & Narrative 1 & Narrative 2 & Narrative 3 & \quad & & Narrative 1 & Narrative 4 & Narrative 5 \\
\midrule
Nar. 1 & 1.00 & 0.95 & 0.96 & & Nar. 1 & 1.00 & 0.92 & 0.93 \\
Nar. 2 & 0.95 & 1.00 & 0.99 & & Nar. 4 & 0.92 & 1.00 & 0.91 \\
Nar. 3 & 0.96 & 0.99 & 1.00 & & Nar. 5 & 0.93 & 0.91 & 1.00 \\
\bottomrule
\end{tabular}
\end{table}

\paragraph{Collection of Pure Narratives}
Rather than studying a single narrative in isolation, this paper is concerned with studying relative prevalence among several narratives. We accordingly introduce the notion of a \emph{collection of pure narratives}, whose representations live in the
embedding space $\mathcal H$. A collection of pure narratives is a collection of narratives that are semantically independent from one another: each retains only an idiosyncratic component, separate from whatever is common across the collection. Purity in this sense is not a property of a single narrative taken in isolation, it is only meaningful relative to the collection it belongs to. In the embedding space, this translate as followed:

\begin{definition}[Collection of pure narratives]
A collection $V_1,\ldots,V_K$ of narratives in $\mathcal H \subset \mathbb R^d$
is a collection of \emph{pure narratives} if the collection $(V_i)_{1 \le i \le K}$ is orthonormal.
\end{definition}

Pure narratives should be understood as an idealized abstraction rather than a text we expect to observe or write down directly: we do not claim that any collection of texts is an exact representation of a collection of pure narratives, and we make no attempt to
produce one. What matters at this stage is only that such an orthonormal collection \emph{exists} in $\mathcal H$; Section~\ref{sec:ortho} then shows how to recover, from any (impure) list of narrative-representing texts, a collection of vectors that is a good practical proxy for it.

In principle a collection of pure narratives for our previous example of (1) \narr{Extreme Heat and Emergency, Policy Response}, (4) \narr{NATO Eastern Flank and the Russia–Ukraine War}, and (5) \narr{IMF Forecasts and Global Growth Slowdown} should have a right panel of Table \ref{tab:cosine_similarity} that is close to a diagonal of ones: they should be pairwise orthogonal (in the cosine similarity sense).

\subsection{A Generative Model on the Embedding Space}
\label{sec:model:generative}

We start by explaining and defining clearly the notion of \emph{prevalence} for pure narratives;
in Section \ref{sec:naive-failure:prevalence} we show how to orthogonalize a family of non-pure narratives to use the same notions on any family of representatives of narratives.

Given a family of pure narratives of interest $(V_i)_{1\le i\le K}$, our goal is to study their relative prevalence within a corpus of texts. We introduce a probabilistic generative model for this purpose, grounded in the linear structure of distributional word vectors~\autocite{arora2016latent} and the Linear Representation Hypothesis in large language models~\autocite{park2024linear}.

\paragraph{Setup.} 
We consider the random variables $(\alpha_1, \ldots, \alpha_K, \varepsilon)$. We place ourselves on the probability space $(\Omega, \cal F, \mathbb P)$ generated by the law of $(\alpha_1, ..., \alpha_K, \varepsilon)$, and equipped with a filtration $\calF_t$. 
The time $\tau(\omega)$ at which an occurrence $\omega \in \Omega$ occurs is a stopping time.
 Conditionally on $\calF_t$, $\alpha_i$ is distributed as a Bernoulli, with a parameter $p_{i,t}$. The process $(p_{i,t})_{t \ge 0}$ is $\calF_t$-adapted. 
 The random variable $\varepsilon$ is such that $\varepsilon = \sqrt{1-\beta^2\textstyle\sum_k\alpha_k}\;u$, with $\beta\in(0,1/\sqrt K\,]$, fixed, and $u$ uniformly distributed on the unit sphere of $\mathcal V^\perp$, independent of $\calF_t$ and $(\alpha_1, \ldots, \alpha_K)$.

We introduce the random variable $X$ and assume that $X(\omega)$ decomposes as
\begin{equation}
    \label{eq:generative}
    X(\omega) = \beta\sum_{i=1}^K \alpha_i(\omega)\, V_i + \varepsilon(\omega).
\end{equation}
Note that $X(\omega)$ is a unit-vector of $\mathbb{R}^d$. 
To keep the notation light, we drop the dependence on $\omega$:
given a database $(\omega_n)_{1 \le n \le N}$ of $N$ occurrences, 
we write $X_n := X(\omega_n)$,
$\alpha_{i,n} := \alpha_i(\omega_n)$, $\varepsilon_n := \varepsilon(\omega_n)$,
and $t_n := \tau(\omega_n)$, so that~\eqref{eq:generative} reads
\begin{equation}
    \label{eq:generative:n}
    X_n = \beta \sum_{i=1}^K \alpha_{i,n}\, V_i + \varepsilon_n, \qquad n = 1,\ldots,N.
\end{equation}

\paragraph{Qualitative interpretation.} If one considers a corpus of texts (for instance a corpus of News) with $N$ occurrences $\omega_n$, the collection $(X_n)_{1 \le n \le N}$ represents the corpus, more precisely, each $X_n$ is the (normalized) vector embedding of a text (for instance of an article). $\tau(\omega_n)=t_n$ is the date at which the article was published. The variable $\alpha_{i,n}$ indicates whether narrative $V_i$ is
present in text $n$. The residual $\varepsilon_n$ captures everything in the text unrelated to the
$K$ narratives of interest: for instance, author style, secondary facts, or unmodeled topics. As it lies in $\mathcal V^\perp$, it carries no component along any narrative of interest.  

\subsection{Narrative Prevalence}
\label{sec:model:prevalence}

\paragraph{Conditional prevalence.}
Recall from Section~\ref{sec:model:generative} that $p_{i,t}$ denotes the
Bernoulli parameter governing the presence of narrative $i$ at time $t$:
$\mathbb E[\alpha_{i} \mid \mathcal F_t] = p_{i,t}$ for any time $t \ge 0$. This quantity is not directly observable, since $\alpha_{i}$ is
itself latent. Only the embedding $X$ is. 

But, we leverage the properties of our model: the orthonormality of the $V_k$ and the fact that $\varepsilon_n \in \text{Vect}(V_1, ..., V_K)^{\perp}$ to obtain, with $X$ defined by (\ref{eq:generative}):
\begin{equation}
    \mathbb E\bigl[V_i^\top X \mid \mathcal F_t\bigr]
    = \beta\sum_{k=1}^K \underbrace{\mathbb{E}\bigl[\alpha_{k} \mid \mathcal F_t\bigr]}_{=p_{k,t}}\, \underbrace{V_i^\top V_k}_{=\,\delta_{ik}} \,
      + \, \mathbb E\bigl[\underbrace{V_i^\top \varepsilon}_{=0} \mid \mathcal F_t\bigr]
    = \beta p_{i,t},
\end{equation}
This shows that, up to a scale, this $p_{i,t}$ coincides with the conditional expectation of the observable
projection $V_i^\top X$, which is what makes it estimable in practice.

\begin{proposition}
\label{prop:prevalence-projection}
The constant $\beta$ is the same for all narratives and all dates and plays no role in comparisons across narratives and time. For simplicity we define $p_i(t) := \beta p_{i,t}$ the \emph{prevalence} (up to a scale) of a narrative in the text corpus at time $t$.  
\begin{equation}
    \label{eq:esp:scalar}
    \mathbb E\bigl[V_i^\top X \mid \mathcal F_t\bigr] = p_i(t).
\end{equation}
\end{proposition}

\begin{definition}[Prevalence of narratives in texts]
\label{def:prevalence}
The previous proposition gives a constructive definition of \emph{the prevalence of narrative $V_i$ in text $X$ appearing in a corpus under the law $d\mathbb P$}, that is 
\begin{equation}
    \label{eq:def:prevalence}
    \hat\alpha_i:=\mathbb E\bigl[V_i^\top X\bigr].
\end{equation}
\end{definition}
This is a well defined notion and can be conditioned by a filtration to get \emph{the prevalence of narrative $V_i$ in a corpus of text at $t$}, that is given by equality (\ref{eq:esp:scalar}).
Note that thanks to our model, the prevalence, that can be read as the contribution of the pure narrative $V_i$ to the text $X$, is nothing more than a scalar product, i.e. a cosine similarity. We show in Section \ref{sec:naive-failure:prevalence} how to extend this estimate to families of non-pure narratives.

\paragraph{Co-occurrence.}
For $i,j \le K$ and any time $t$, we define:
\begin{equation}
    \label{eq:cooccurrence}
    p_{ij,t} := \mathbb E\bigl[\alpha_{i}\alpha_{j} \mid \mathcal F_t\bigr],
\end{equation}
the conditional probability that an article published at $t$ carries both
narratives $i$ and $j$ simultaneously; note $p_{ii,t}=p_{i,t}$. It is not directly observable, since
$\alpha_{i},\alpha_{j}$ are themselves latent. The following proposition
shows it is recovered exactly from the
observable product of projections $(V_i^\top X)(V_j^\top X)$.

\begin{proposition}
\label{prop:cooccurrence-projection}
For any time $t \ge 0$,
\begin{equation}
    (V_i^\top X)(V_j^\top X) = \beta^2\alpha_{i}\alpha_{j} \quad,
    \qquad\text{and hence}\qquad
    \mathbb E\bigl[(V_i^\top X)(V_j^\top X)\mid \mathcal F_t\bigr] = \beta^2 p_{ij,t}.
\end{equation}
As in ~\ref{prop:prevalence-projection}, $\beta^2$ plays no role in comparisons so for simplicity, we define $p_{ij}(t) = \beta^2 p_{ij,t}$, the \emph{co-occurrence} of narratives $i$ and $j$ at time $t$. Up to a scale, it corresponds to the conditional probability that a text published at $t$ carries both narratives $i$ and $j$ simultaneously. So finally we have: 
\begin{equation}
    E\bigl[(V_i^\top X)(V_j^\top X)\mid \mathcal F_t\bigr] =  p_{ij}(t)
\end{equation}
\end{proposition}

\subsection{Empirical Estimators}
\label{sec:model:estimator}

\paragraph{Time discretization}
Time is discretized into periods indexed by $t \in \{1,\ldots,T\}$,
corresponding to the values taken by the stopping time $\tau$ once the
observation window is partitioned into days, weeks, or months, without fixing
\emph{a priori} the granularity. For each period $t$, we denote by
\begin{equation}
    \mathcal C_t := \{n : \tau(\omega_n) = t\}, \qquad n_t := |\mathcal C_t|
\end{equation}
the set and number of articles published during period $t$. Since $\tau$ is
an $(\mathcal F_t)$-stopping time, $\{\tau(\omega_n) = t\}$ is $\mathcal
F_t$-measurable for each $n$, and hence so are $\mathcal C_t$ and $n_t$.

This naturally leads to unbiased empirical estimators for the prevalence, and the co-occurrence.
\begin{align}
    \hat p_i(t) &= \frac{1}{n_t} \sum_{n \in \mathcal C_t} V_i^\top X_n,
    \label{eq:estimator_p}\\
    \hat p_{ij}(t) &= \frac{1}{n_t} \sum_{n\in\mathcal C_t} (V_i^\top X_n)(V_j^\top X_n).
\end{align}
All proofs are detailed in \ref{app:proofs}.

\section{Failure of the Natural Approach: why orthonormality matters}
\label{sec:naive-failure}

In Section \ref{sec:narratives:embedding} we made the difference between \emph{representatives of narratives}, that are sentences describing in human language the targeted narrative, 
and 
\emph{collections of pure narratives}, namely set of orthogonal narratives.
In Section \ref{sec:model:generative} we have shown that it is straightforward to build unbiased estimates of the \emph{prevalence of narratives in a corpus of texts}.

It would have been natural to take as vector-representatives for our targeted narratives, the embedding obtained by encoding text-representatives through a language model. Nevertheless we have illustrated in the same Section \ref{sec:narratives:embedding} that  such representatives of different narratives (we took the example of (1) \narr{Extreme heat}, (4) \narr{Russia-Ukraine war} and (5) \narr{IMF Forecasts}) are not at all orthogonal: they have a lot in common due to the English structure and the common semantic used to describe clearly any narrative in one short paragraph.

This section show why not using a collection of pure narratives would fail at estimating the prevalence and other quantities of interest. Not having access to a pure version of the collection of narratives prevent to simply estimate the importance (or joined importance) of narratives of interest in texts.

The theoretical framework developed in
Section~\ref{sec:model} lets us establish precisely why this natural approach fails. We quantify it both at the level of
prevalence (a first-moment quantity, Section~\ref{sec:naive-failure:prevalence})
and at the level of co-occurrence
(Section~\ref{sec:naive-failure:cooccurrence}).

\subsection{Prevalence}
\label{sec:naive-failure:prevalence}

Let $(N_i)_{1\le i \le K} \in \mathcal{H}$ be a family of unit-vector embedding that represents our studied narratives. The family is not assumed to be orthogonal. Analogously to Section~\ref{sec:model:prevalence}, the
{natural} approach measures the prevalence of $N_i$ at time $t$ as
\begin{equation}
    \label{eq:naive-prevalence}
    \pi_{N_i}(t) := \mathbb E\bigl[N_i^\top X \mid \mathcal F_t\bigr]
\end{equation}
\medskip

Since $N_i$ and $X$ are both unit-norm, $N_i^\top X_n$ is exactly their
cosine similarity; $\pi_{N_i}(t)$ is then the average cosine similarity
between narrative $i$'s candidate embedding and the articles published
during period $t$, precisely the quantity the natural approach would compute in practice.

Since $(V_k)_{k\le K}$ is an orthonormal basis of $\mathcal V$, $N_i$
decomposes uniquely as
\begin{equation}
    \label{eq:candidate-decomp}
    N_i = \sum_{k=1}^K c_{ik}\, V_k + N_i^\perp, \qquad c_{ik} := N_i^\top V_k, \quad N_i^\perp \in \mathcal V^\perp.
\end{equation}

\begin{proposition}[Naive prevalence under an impure candidate]
\label{prop:naive-prevalence}
\begin{equation}
    \pi_{N_i}(t) = c_{ii}\, p_i(t) + \sum_{k\ne i} \underbrace{c_{ik}\, p_k(t)}_{\text{leakage}}.
\end{equation}
\end{proposition}

\paragraph{Interpretation.}
When $(N_i)_i$ is pure, $c_{ii}=1$ and $c_{ik}=0$ for $k\ne i$, and
$\pi_{N_i}(t)=p_i(t)$ exactly, recovering the unbiased estimator of
Proposition~\ref{prop:prevalence-projection}. For an impure candidate,
however, $\pi_{N_i}(t)$ mixes the true prevalence $p_i(t)$ with every other
narrative's prevalence $p_k(t)$, weighted by the leakage coefficient
$c_{ik}=N_i^\top V_k$. If a historical event drives up the prevalence of an
unrelated but semantically connected narrative $k$ ($p_k(t)$ increases), the
naive indicator $\pi_{N_i}(t)$ increases mechanically, creating the illusion
that narrative $i$ itself has become more prevalent.

\subsection{Co-occurrence of Two Narratives}
\label{sec:naive-failure:cooccurrence}

Recall from Section~\ref{sec:model:prevalence} that the true co-occurrence of
narratives $i$ and $j$ at time $t$, $p_{ij}(t)$ coincides exactly with $\mathbb
E[(V_i^\top X)(V_j^\top X)\mid\mathcal F_t]$ for pure narratives
(Proposition~\ref{prop:cooccurrence-projection}). For arbitrary candidates
$W_i,W_j\in\mathcal H$, the naive approach measures their co-occurrence by
transplanting this same estimator to the candidates it actually has access
to \textit{i.e} language-model embedding of a textual description of each narrative,
rather than the (unobservable) pure narratives themselves. Concretely, for
each article $n$, $X_n^\top W_i$ is the cosine similarity between the article
and candidate $i$; their product is large only when the article aligns
strongly with \emph{both} candidates at once, and small otherwise, so that
averaging this product over the pool of articles published at $t$ is,
intuitively, a natural proxy for how often (and how strongly) narratives $i$
and $j$ jointly manifest in the corpus. We thus define the following metric:
\begin{equation}
    \label{eq:naive-cooccurrence}
    M_{ij}(t) := \mathbb E\bigl[(X^\top W_i)(X^\top W_j)\mid\mathcal F_t\bigr].
\end{equation}

\begin{proposition}[Naive co-occurrence under impure candidates]
\label{prop:naive-cooccurrence}
Write $W_i = \sum_k c_{ik}V_k + \eta_i$ and $W_j = \sum_l c_{jl}V_l+\eta_j$ as
in~\eqref{eq:candidate-decomp}. Then for $i \neq j$,
\begin{equation}
    \label{eq:naive-cooccurrence-decomp}
    M_{ij}(t) = (c_{ii}c_{jj} + c_{ij}c_{ji})p_{ij}(t) +  \underbrace{\sum_{(k,l)\ne (i,j), (j,i)} c_{ik}\,c_{jl}\,p_{kl}(t)}_{\text{narrative leakage}} \;+\; \underbrace{\sigma_t^2\eta_i^\top\,\eta_j}_{\text{noise term}}, 
\end{equation}
where $\sigma_t^2 = \dfrac{1- \beta^2 \sum_k p_{k,t}}{d-K}$.
\end{proposition}

\paragraph{Pure candidates.}
When $W_i=V_i$ and $W_j=V_j$, $c_{ik}=\delta_{ik}$, $c_{jl}=\delta_{jl}$, and
$\eta_i=\eta_j=0$: both the signal term collapses to $p_{ij}(t)$ and the
noise term vanishes identically, so $M_{ij}(t)=p_{ij}(t)$
exactly. This is just what we had in Proposition~\ref{prop:cooccurrence-projection}.

\paragraph{Impure candidates}
When $W_i,W_j$ are not pure, \eqref{eq:naive-cooccurrence-decomp} shows that
$M_{ij}(t)$ departs from $p_{ij}(t)$ through two distinct mechanisms, each
sufficient on its own to bias the naive measure.
\begin{itemize}
    \item \emph{Narrative leakage}.
    The signal term mixes \emph{every} entry of the true
    co-occurrence matrix $(p_{kl}(t))_{k,l}$, not just $p_{ij}(t)$ itself. In
    particular, even if narratives $i$ and $j$ never truly co-occur
    ($p_{ij}(t)=0$), $M_{ij}(t)$ can be large  because $W_i$ and $W_j$ both
    carry a nonzero loading on a shared third narrative $k$ ($c_{ik}\ne0$ and
    $c_{jk}\ne0$): the naive approach then reports an artificial
    synchronization between two narratives that share nothing in the corpus.

    \item \emph{Semantic Noise}.
    The noise term, $\sigma_t^2 \eta_i^\top \eta_j$,
    is a second source of artificial co-variation. Even a candidate pair with
    perfectly targeted signal ($c_{ik}=\delta_{ik}$, $c_{jl}=\delta_{jl}$, zero
    leakage onto any other narrative) still registers a spuriously elevated
    $M_{ij}(t)$ as soon as their residuals $\eta_i,\eta_j$ 
    (whatever in each candidate's textual representation is unrelated to any
    narrative, e.g.\ shared vocabulary or phrasing used to describe both) happen
    to be non-orthogonal. Since $\sigma_t^2>0$, any $\eta_i^\top\eta_j>0$
    mechanically inflates $M_{ij}(t)$ above the true co-occurrence, regardless of
    whether the narratives themselves ever co-occur in the corpus: two narratives
    can appear artificially correlated purely because their candidate
    descriptions happen to be phrased in a similar style.
\end{itemize}

The natural approach therefore fails on two distinct counts. At the level of
prevalence, it conflates a narrative's true presence with leakage from every
other narrative sharing representational overlap with it
(Proposition~\ref{prop:naive-prevalence}). At the level of co-occurrence, it
conflates genuine joint presence with two sources of spurious
correlation: leakage onto narratives, and alignment of narrative-unrelated residuals, induced purely by the geometry of the
candidate representations (Proposition~\ref{prop:naive-cooccurrence}). Both
failure modes vanish exactly, once the candidates are
replaced by their pure (orthogonalized) counterparts which is precisely
what the procedure of Section~\ref{sec:ortho} constructs.

\section{Proposed Methodology: from Narratives to Quantitative Monitoring of their Relative Prevalence}
\label{sec:method}

In this section, we detail the  methodology to monitor the relative prevalence of narratives. 

\paragraph{Step 1: Choose the narratives.}
A narrative is an abstract idea, claim, or framing about the world
(Section~\ref{sec:narratives:embedding}). The collection to monitor is therefore
a conceptual choice made \emph{a priori} by the analyst, driven by the question
asked (for instance, which competing stories are told about a given
phenomenon?) rather than by the corpus. Contrary to topic models, a narrative
need not be statistically prevalent in the corpus to be measured, and, contrary
to supervised classifiers, no labeled examples are required.

\paragraph{Step 2: Vectorize the narratives.}
An idea cannot be passed to an encoder, so each narrative $i$ is represented by
a short descriptive text $T_i$ of one to three sentences. The descriptions
should be homogeneous in length, register, and style, so that the encoder
reflects differences of content rather than differences of phrasing. Each
$T_i$ is embedded with a text encoder and normalized, yielding the
\emph{raw narrative vectors}
\[
  N_i^{*} = \frac{f(T_i)}{||f(T_i)||} \in \R^d, \qquad i=1,\ldots,K,
\]
where $f$ denotes the encoder (here \emph{ModernBERT}, $d=768$). In our
experiments the descriptions were drafted with the assistance of a large
language model to ensure stylistic homogeneity (\ref{app:narratives}).

\paragraph{Step 3: Construct the pure narratives.}
The raw vectors $(N_i^*)_i$ are not pure: texts written in the same language,
with the same journalistic register, share most of their embedding. We
therefore solve the max-min problem~\eqref{eq:maxmin} on the Stiefel manifold
(Section~\ref{sec:ortho:stiefel}) and obtain an orthonormal family
$(V_i)_{i\le K}$, in which each $V_i$ is as close as possible to $N_i^*$
(in the worst case over $i$) and orthogonal to all the others. These vectors
are the practical proxies for the collection of pure narratives of
Section~\ref{sec:model}.

\paragraph{Illustration.}
Consider two narratives of \ref{app:narratives_illustration}:
\narr{Extreme Heat and Emergency Policy Response}, which asserts that
record-breaking temperatures force governments into emergency action, and
\narr{International Climate Commitments Under Pressure}, which asserts that
states are failing to honor the Paris Agreement. These are distinct claims, but
their textual representatives share a substantial set of concepts: climate
change and global warming, greenhouse gas emissions, governments and public
policy. They also share the generic register of a journalistic English
sentence, which is why even narratives on unrelated topics have raw cosine
similarities above $0.9$ (Table~\ref{tab:cosine_similarity}). Consequently, an
article about an upcoming climate summit aligns strongly with \emph{both} raw
vectors, even if it does not mention heatwaves, and the two scores rise and
fall together whenever climate is in the news. This is precisely the leakage
of Proposition~\ref{prop:naive-prevalence}: the measured prevalence of one
narrative is contaminated by the prevalence of every narrative with which it
shares content. After orthogonalization, the shared component is removed from
each vector. Ideally, what remains of the first narrative is the link between an
extreme weather event and an emergency policy response, and what remains of the
second is adherence to international commitments, so that each score now
responds only to its own idea. 

\paragraph{Step 4: Vectorize the corpus.}
Each text $n$ of the corpus is embedded with the
\emph{same} encoder $f$ and normalized, giving $X_n$, and is stored with its
publication date $t_n$. This yields the database $(t_n,X_n)_{1\le n\le N}$ of
Section~\ref{sec:model:generative}. Using the same encoder is necessary for the
narrative vectors and the texts to live in the same space $\mathcal H$.

\paragraph{Step 5: Estimate prevalence.}
Time is discretized into periods (here, weeks), and for each narrative $i$ and
period $t$ we compute
\[
  \hat p_i(t) = \frac{1}{n_t}\sum_{n\in\mathcal C_t} V_i^\top X_n ,
\]
the average cosine similarity between the pure narrative and the texts of
the period. By Proposition~\ref{prop:prevalence-projection}, $\hat p_i(t)$ is
an unbiased estimator of the prevalence of narrative $i$. Co-occurrences $\hat p_{ij}(t)$
are obtained in the same way, and the resulting series can be smoothed for
visualization or tested for ruptures, as in
Section~\ref{sec:application:quantitative}.

\paragraph{Practical remarks.}
First, purity is relative to the collection: each $V_i$ depends on all the
other narratives, so adding, removing, or rewriting one of them requires
recomputing the whole family, and prevalences are comparable only within a
given collection. Second, orthonormality requires $K\le d$, which is
immaterial in practice. Third, the collection of texts should not contained two distinct texts corresponding to the same conceptual narrative (eg. \narr{Trump has won the presidential elections} and \narr{After the elections, the new president is Donald Trump}).

\begin{figure}[H]
    \centering
    \includegraphics[scale=.7]{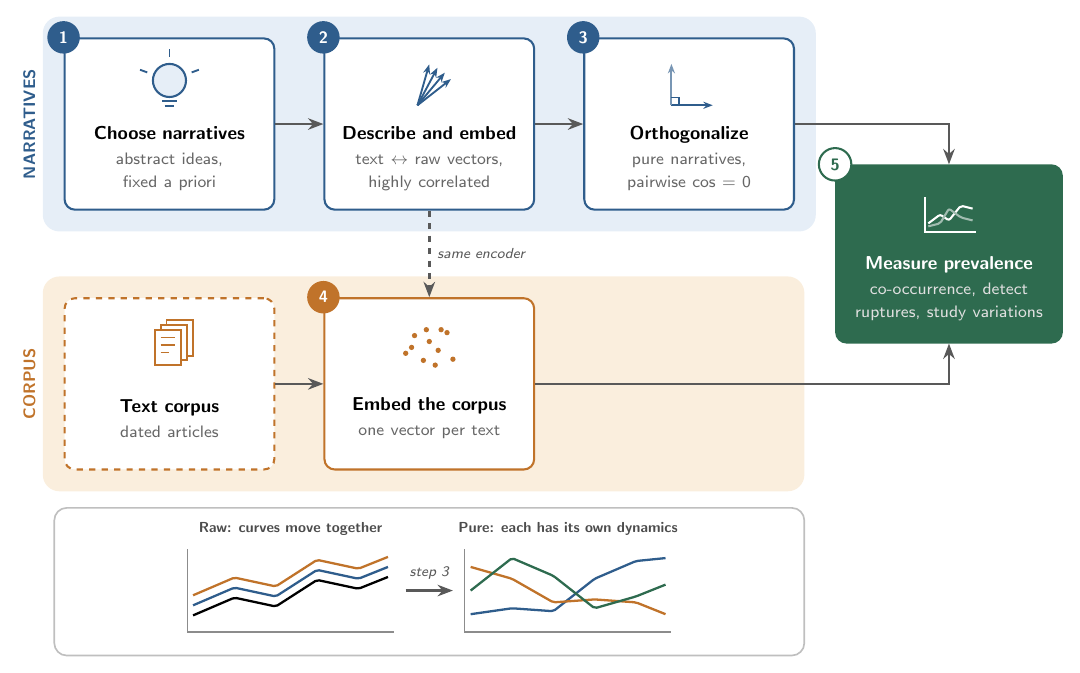}
    \caption{Pipeline to monitor the prevalence of a collection of narratives.
Narratives are chosen conceptually, described in text, embedded, and
orthogonalized into pure narratives. The corpus is embedded with the same
encoder, and the two branches meet only in the final step, where prevalence
is measured. The inset illustrates the effect of step~3: raw
narrative curves move almost synchronously, whereas the curves of pure
narratives follow their own dynamics (schematic).}
    \label{fig:my_label}
\end{figure}

\section{Orthogonalization of Narratives}
\label{sec:ortho}

The previous section demonstrates that \emph{arbitrary representatives of narratives do not have enough good properties to enable to estimate their prevalence in texts}.
Nevertheless the modelling angle we exposed in Section \ref{sec:model:generative} allows to \emph{establish an orthonormalization procedure allowing to recover a nice candidate collection of pure narratives}.
It will be nice in the sense: it faithfully represents the collection of narratives spanned by chosen representatives, and it is pure.

\subsection{The Problem: Correlated Narrative Vectors}
\label{sec:ortho:problem}

Let $\{N_i\}_{i=1}^{n} \subset \R^d$ denote the embedding vectors of $n$ narratives,
obtained by passing their textual 
representatives through a language model.
A fundamental difficulty is that narratives written in the same language,
sharing a journalistic style and treating related themes, yield representations
that are highly correlated in the embedding space.
In our experiment (Section~\ref{sec:narratives:embedding}), pairwise cosine similarities
between raw narrative vectors lie in the range $[0.91, 0.99]$.

A direct consequence is that the time series of cosine similarities
$\cos(N_i, X_t)$ (where $X_t$ is the embedding of an article published at time $t$)
evolve almost synchronously across narratives, reflecting global topic volume
rather than the relative dominance of specific narratives.

The goal is therefore to construct representative vectors $\{V_i\}_{i=1}^{n}$
that satisfy:
\begin{itemize}
    \item \textbf{Faithfulness}: each $V_i$ is as close as possible to $N_i^* = N_i / ||N_i||$;
    \item \textbf{Orthonormality}: $V_i^\top V_j = \delta_{ij}$.
\end{itemize}

\subsection{Monotonous Orthogonalization and Its Limits}
\label{sec:ortho:gram}

A natural candidate is the Gram--Schmidt algorithm.
However, Gram--Schmidt is strongly order-dependent: the first vector is unchanged
while subsequent vectors undergo increasingly large deformations, introducing an
arbitrary bias.
Moreover, it does not optimize any global fidelity criterion.
Importantly, because the procedure is not incremental, any addition or removal of a
narrative requires recomputing the entire orthogonal basis to guarantee
optimality and order-independence. 

\subsection{Global Procrustes Alignment and Its Limits}
\label{sec:ortho:procrustes}

A natural alternative is the Orthogonal Procrustes solution {\autocite{schonemann1966generalized}}, which finds the matrix $V$ that minimizes the global Frobenius norm $||N^* - V||_F^2$. While this admits an exact closed-form solution via Singular Value Decomposition (SVD), it maximizes the \emph{sum} of alignments rather than the \emph{minimum}. In the context of highly correlated narratives, Procrustes optimization can sacrifice the representation of minority narratives: it allows individual vectors $V_i$ to be pushed far from $N_i^*$ if doing so slightly improves the overall global alignment. To prevent any single narrative from being discarded or severely distorted, we instead require a max-min formulation.

\subsection{Global Optimization on the Stiefel Manifold}
\label{sec:ortho:stiefel}

To overcome these limitations, we seek the orthonormal family
$\{V_i\}_{i=1}^{n}$ that maximizes the minimum alignment with the original narratives,
i.e.\ we solve the max-min problem:
\begin{equation}
    \label{eq:maxmin}
    \underset{(V_1,\ldots,V_n)}{\arg\max}\;
    \min_{i=1,\ldots,n}\; N_i^{*\top} V_i
    \qquad
    \text{subject to}
    \quad
    \begin{cases}
        ||V_i|| = 1 & \forall\, i,\\
        N_i^{*\top} V_i \geq 0 & \forall\, i,\\
        V_i^\top V_j = 0 & \forall\, i \neq j.
    \end{cases}
\end{equation}
The constraint $N_i^{*\top} V_i \geq 0$ ensures that $V_i$ and $N_i^*$
are on the same side of the hyperplane, i.e.\ that $V_i$ is a genuine representative
of $N_i$ rather than its antipode.
The max-min objective guarantees that the algorithm focuses on the worst-case
narrative, preventing any single narrative from being sacrificed.

Stacking the row vectors $V_i^\top$ into a matrix $V \in \R^{n \times d}$ (and $N_i^{*\top}$ into $N^* \in \R^{n \times d}$),
the orthonormality constraints are equivalent to $V V^\top = I_n$,
i.e.\ $V$ belongs to the \emph{Stiefel manifold}~\autocite{boumal2023optimization,wen2013feasible}:
\begin{equation}
    \label{eq:stiefel-manifold}
    \St(n,d) = \bigl\{V \in \R^{n \times d} : V V^\top = I_n\bigr\}.
\end{equation}

Because the objective $\min_{1 \le i \le n}(N_i^{*\top} V_i)$ in~\eqref{eq:maxmin} is non-differentiable,
we approximate it with the smooth \emph{softmin} surrogate parameterized by temperature $\tau > 0$:
\begin{equation}
    \label{eq:softmin}
    f(V) = \tau \log \sum_{i=1}^{n} \exp\!\left(-\frac{N_i^{*\top} V_i}{\tau}\right).
\end{equation}
Minimizing $f(V)$ on $\St(n,d)$ is performed via \emph{Riemannian gradient descent},
with three steps per iteration:
\begin{enumerate}
    \item \textbf{Euclidean gradient.}
    \begin{equation}
        \nabla f(V) = -\operatorname{diag}(w)\, N^*,
        \qquad
        w_i = \frac{\exp(-N_i^{*\top} V_i / \tau)}{\sum_{j=1}^{n} \exp(-N_j^{*\top} V_j / \tau)}.
    \end{equation}

    \item \textbf{Riemannian gradient.}
    Project onto the tangent space $T_V \St(n,d) = \{\Delta \in \R^{n \times d} : \Delta V^\top + V \Delta^\top = 0\}$:
    \begin{equation}
        \operatorname{grad} f(V)
        = \nabla f(V) - \sym\!\bigl(\nabla f(V)\, V^\top\bigr)\, V,
        \qquad
        \sym(A) = \tfrac{1}{2}(A + A^\top).
    \end{equation}

    \item \textbf{Retraction.}
    After the descent step $\tilde{V} = V - \alpha\,\operatorname{grad} f(V)$, the updated matrix $\tilde{V} \in \R^{n \times d}$ no longer satisfies $\tilde{V} \tilde{V}^\top = I_n$ in general. To map $\tilde{V}$ back onto $\St(n,d)$~\autocite[Chapter~4]{absil2008optimization}, we compute the thin QR factorization of its transpose, $\tilde{V}^\top = Q R$, where $Q \in \R^{d \times n}$ has orthonormal columns ($Q^\top Q = I_n$) and $R \in \R^{n \times n}$ is upper triangular with $\operatorname{diag}(R) > 0$. The retracted point is then defined as $V_{\mathrm{new}} = Q^\top \in \St(n,d)$.
\end{enumerate}
Initialized at the symmetric Procrustes solution $V^{(0)} = (N^* N^{*\top})^{-1/2} N^*$, this iterative scheme guarantees that the row-orthonormality constraints are satisfied at every iteration while converging to a local optimum of~\eqref{eq:softmin}.

\subsection{Interpretation of the Orthogonalized Vectors}
\label{sec:ortho:interpretation}

\begin{figure}[hbt!]
    \centering
    \includegraphics[width=\linewidth]{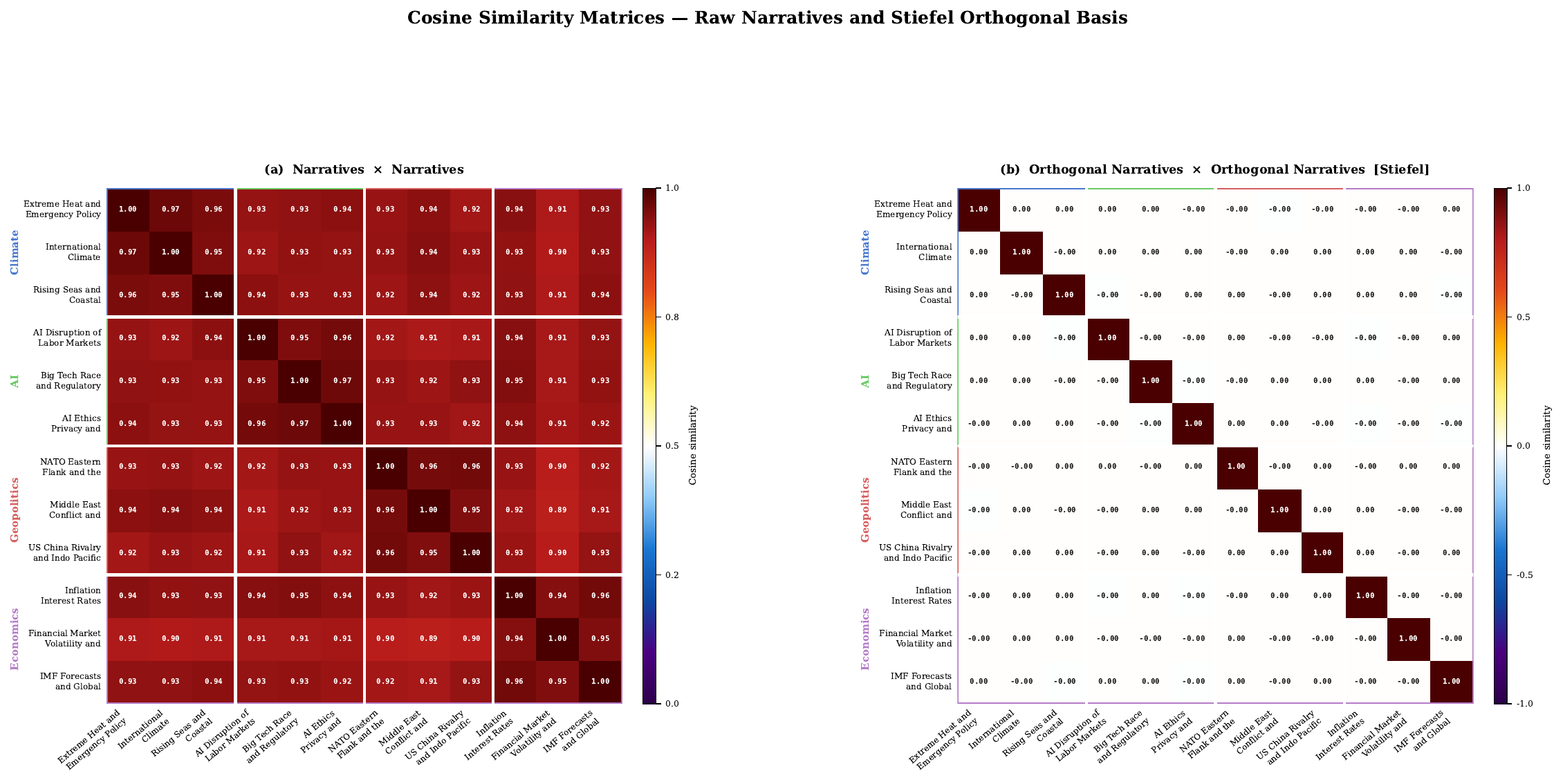}
  \caption{Cosine similarity matrices of narrative embedding before and after Stiefel orthogonalization. Panel~(a) shows the raw \textit{ModernBERT} embedding, where diagonal blocks are systematically darker, reflecting stronger intra-chapter semantic 
cohesion. Panel~(b) shows the orthogonalized embedding after Stiefel optimization, yielding zero pairwise cosine similarity across all narrative pairs thereby isolating the idiosyncratic semantic signal 
of each narrative.}
    \label{fig:cosine_matrix1}
\end{figure}

\begin{figure}[hbt!]
    \centering
    \includegraphics[width=\linewidth]{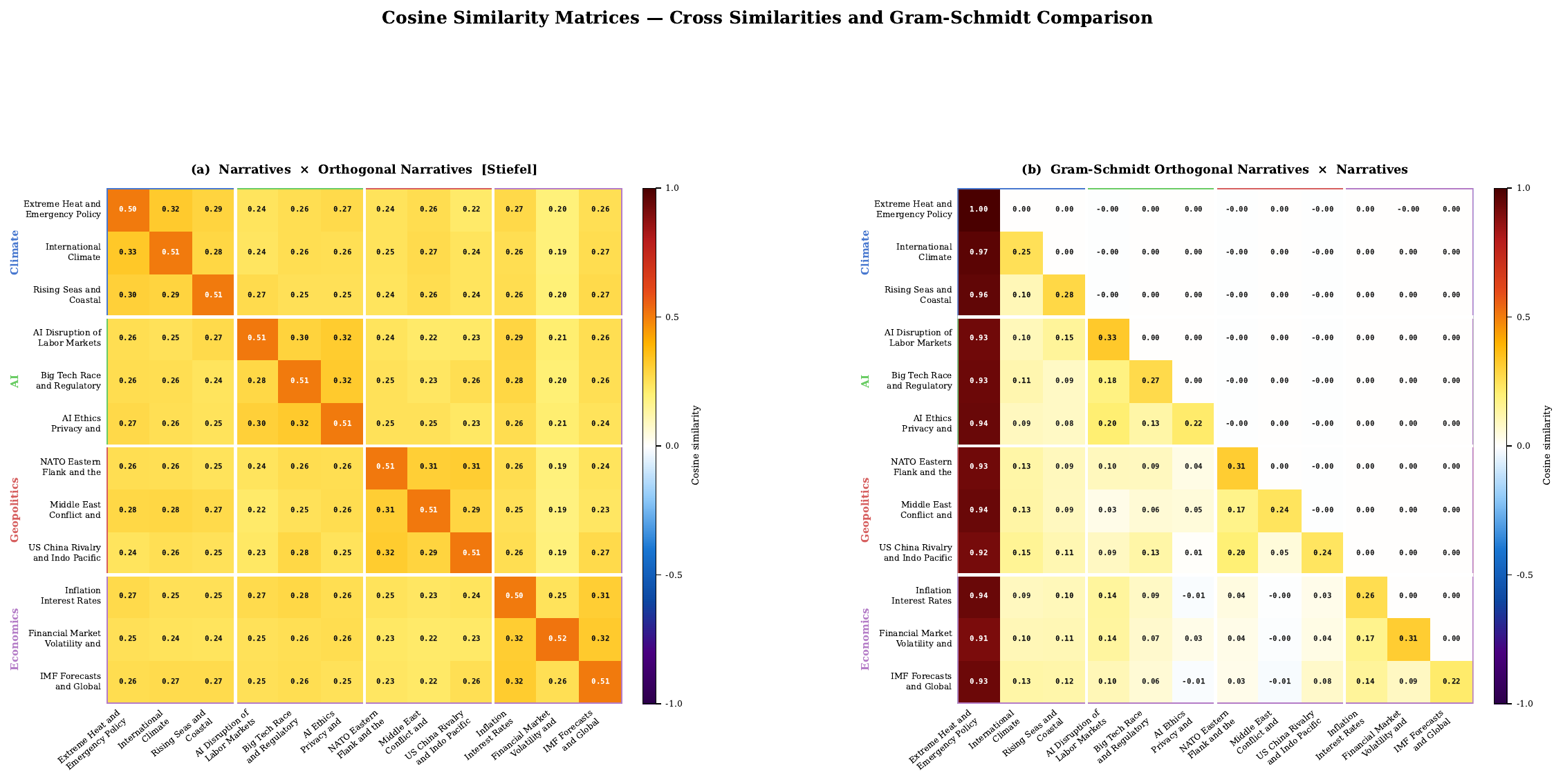}
\caption{Cross-similarity matrices between raw narrative embedding and their orthogonalized representatives. Panel~(a) shows the Stiefel-optimized representatives, where a clear diagonal structure confirms that each representative retains the idiosyncratic semantic content of its source narrative. Panel~(b) shows the 
Gram-Schmidt representatives, revealing a severe order-dependent distortion: the first narrative is left unchanged and remains correlated with all others, to the point where its representative exhibits higher similarity with unrelated raw narratives than some narratives' own representatives do with themselves.}
    \label{fig:cosine_matrix2}
\end{figure}

We first evaluate the geometric behavior of our orthogonalization procedure on the twelve narratives of the Contemporary Policy Collection (Appendix~\ref{app:narratives_illustration}, Table~\ref{tab:narratives}), which span four modern macro-thematic chapters (\emph{Climate}, \emph{Artificial Intelligence}, \emph{Geopolitics}, and \emph{Economics}). The four panels of Figure~\ref{fig:cosine_matrix1} and Figure~\ref{fig:cosine_matrix2} illustrate the behavior of the original narrative embeddings and their orthogonalized counterparts under both the Stiefel-based procedure and the classical Gram--Schmidt algorithm.

Panel (a) of Figure ~\ref{fig:cosine_matrix1} displays the cosine similarity matrix between the raw narrative embedding. A striking feature is the overall high level of correlation across all narrative pairs, reflecting the anisotropy of the embedding space produced by \textit{ModernBERT}. Diagonal blocks are systematically darker, confirming that narratives belonging to the same thematic chapter are more strongly correlated with one another than with narratives from other chapters. For instance, \textit{"AI Disruption of Labor Markets"} and \textit{"Big Tech Race and Regulatory Pushback"} exhibit a cosine similarity of 0.95, whereas the similarity between \textit{"AI Disruption of Labor Markets"} and \textit{"Financial Market Volatility and Fed Signaling"} is lower (with cosine similarity of 0.91). This intra-chapter cohesion validates the semantic consistency of the narrative groupings. 
The text-representative narratives taken as examples in Section \ref{sec:narratives:embedding} come from this list (see Appendix~\ref{app:narratives_illustration}).

Panel (b) of Figure ~\ref{fig:cosine_matrix1}  demonstrates the success of the Stiefel optimization. The resulting representative matrix is fully orthogonal, with pairwise cosine similarities equal to zero across all narrative pairs. Crucially, this orthogonality holds not only across thematic chapters but also within chapters: \textit{"Extreme Heat and Emergency Policy Response"} is equally orthogonal to \textit{"International Climate Commitments Under Pressure"} and \textit{"US China Rivalry and Indo Pacific"}. The procedure thus isolates the idiosyncratic component of each narrative, stripping away both cross-chapter and intra-chapter shared variance. What remains is the semantic signal that is strictly proper to each narrative, independently of all others.

Panel (a) of Figure ~\ref{fig:cosine_matrix2} presents the cross-similarity between raw narratives and their Stiefel-optimized representatives. A clear diagonal structure emerges: each representative is more strongly correlated with its own source narrative than with any other, confirming that the optimization procedure preserves the idiosyncratic semantic content of each narrative while beginning to disentangle inter-narrative correlations.

Panel (b) of Figure ~\ref{fig:cosine_matrix2} exposes the limitations of the classical Gram-Schmidt orthogonalization. This procedure is well known to be sensitive to the ordering of input vectors: by construction, the first vector in the sequence is left entirely unchanged, while subsequent vectors are progressively deflated. As a result, the first representative, here that of \narr{"Extreme Heat and Emergency Policy Response"}, retains its original embedding and remains correlated with all other raw narratives. The pathological consequence is particularly striking: the representative of \narr{"Extreme Heat and Emergency Policy Response"} exhibits a higher cosine similarity with the raw narrative \narr{"Middle East Conflict and Humanitarian Crisis"}, for instance, than the representative of \narr{"Middle East Conflict and Humanitarian Crisis"} does with the very narrative it is supposed to represent. This severe distortion renders the Gram-Schmidt basis semantically uninterpretable as a set of narrative representatives, and underscores the superiority of the Stiefel-based approach for this task.

\subsection{Why Not Pre-Center? Anisotropy, Monotonicity, and the Stiefel Trade-Off}
\label{sec:ortho:tradeoff}

Contextualized language encoders map natural-language sentences into a narrow
\emph{anisotropic cone} in $\R^d$, so that even semantically unrelated sentences
share high positive cosine similarity along a common linguistic background axis
$m \in \mathcal{V}^\perp$ ($\norm{m}=1$)~\autocite{arora2016latent,ethayarajh2019contextual}.
In the representation-learning literature, a standard post-processing heuristic is
\emph{``All-but-the-Top''}~\autocite{mu2018all}, which subtracts the empirical mean
vector (and top principal directions) before renormalizing onto the unit sphere.
It is therefore natural to ask why we orthogonalize the un-centered candidate
embeddings $(N_1^*,\ldots,N_K^*)$ directly on $\St(K,d)$ rather than centering first.

Under the canonical equi-correlated cone decomposition
$N_i^* = \sqrt{1-\rho}\,V_i + \sqrt{\rho}\,m$ ($\rho \in (0,1)$) and anisotropic
residual $\varepsilon = \Gamma m + \tilde{\varepsilon}$ ($\E[\Gamma \mid \calF_t] = \gamma_t \in (0,1)$),
three exact geometric results govern this choice:
\begin{enumerate}
    \item \textbf{Remaining ${\cal O}(1/\sqrt{K})$ cone residual without centering.}
    The global max-min Stiefel solution~\eqref{eq:maxmin} admits the closed form
    $\hat{V}_i = V_i - \frac{1-\lambda}{K}\sum_{k=1}^K V_k + \mu m$---coinciding with the
    symmetric orthogonalization of~\textcite{lowdin1950nonorthogonality}---where
    $\lambda = \sqrt{(1-\rho)/(1+(K-1)\rho)}$ and
    $\mu = \sqrt{\rho/(1+(K-1)\rho)} = {\cal O}(1/\sqrt{K})$.
    Consequently, un-centered Stiefel projection damps---but does not completely
    annihilate---the shared cone, leaving an ${\cal O}(1/\sqrt{K})$ intercept $\mu\gamma_t$
    in prevalence ($\approx 0.26$ at $K=12,\rho=0.93$) and an ${\cal O}(1/K)$ floor
    $\mu^2\E[\Gamma^2\mid\calF_t]$ in pairwise co-occurrence ($\approx 0.071$ in
    Table~\ref{tab:covariation_stats}), alongside an ${\cal O}(1/K)$ demeaned shift
    $-\frac{1-\lambda}{K}\sum_{k=1}^K p_k(t)$.

    \item \textbf{Loss of spherical monotonicity and rank under centering.}
    We deliberately refrain from centering to preserve \emph{strict monotonicity} of
    narrative scoring. Along the unit-sphere arc $X(\theta) = \sin\theta\,V_i + \cos\theta\,m$
    ($\theta \in [0,\pi/2]$), centering by $\gamma m$ and renormalizing onto $\mathbb{S}^{d-1}$
    via $\tilde{X}(\theta) = (X(\theta)-\gamma m)/\norm{X(\theta)-\gamma m}$ yields a cosine
    score $f(\theta) = V_i^\top \tilde{X}(\theta)$ whose derivative
    \[
      f'(\theta) = \frac{(\cos\theta - \gamma)(1 - \gamma\cos\theta)}{\bigl(\sin^2\theta + (\cos\theta - \gamma)^2\bigr)^{3/2}}
    \]
    turns \emph{strictly negative} for all $\theta \in (\arccos\gamma, \pi/2]$, penalizing
    articles most strongly aligned with $V_i$ (reversing above $\approx 25.8^\circ$ at
    $\gamma = 0.90$), while empirical corpus centering $\bar{X}$ additionally contaminates
    directions with dominant corpus narratives and introduces look-ahead dependence.
    Conversely, centering only the $K$ candidates ($N_i^* - \bar{N}^*$) collapses their span
    to rank $K - 1$, precluding $K$ orthonormal representatives. By contrast, un-centered
    Stiefel scoring $\hat{V}_i^\top X$ is globally strictly increasing in $\alpha_i$ under~\eqref{eq:generative}
    and strictly increasing on $X(\theta)$ up to $\arctan\bigl((1-(1-\lambda)/K)/\mu\bigr) \approx 72.7^\circ$,
    strictly encompassing the empirical embedding cone.

    \item \textbf{Impossibility trilemma and minimax optimality of Stiefel.}
    More generally, no corpus-free orthonormal basis $W \in \St(K,d)$ of
    $\mathrm{Vect}(N_1^*,\ldots,N_K^*)$ (with $\bar{N}^{*\top}\bar{W} \ge 0$) paired with spherical
    centering $\psi_c(X) = (X - cm)/\norm{X - cm}$ ($c \in [0,1)$) can simultaneously achieve
    full-rank orthonormality, zero baseline on narrative-free cone texts, and global spherical
    monotonicity on $[0,\pi/2]$. Within $\St(K,d)$, Parseval's identity enforces
    $\sum_{i=1}^K (W_i^\top m)^2 = K\mu^2$, so $\max_{1\le i\le K} |W_i^\top m| \ge \mu = K^{-1/2}\bigl(1 + {\cal O}(K^{-1})\bigr)$
    for \emph{every} orthonormal basis of the candidate span. Whereas sequential Gram--Schmidt
    concentrates an ${\cal O}(1)$ cone loading $|W_1^{\mathrm{GS}\top} m| = \sqrt{\rho}$ (independent of $K$) onto the first vector,
    the Stiefel solution $\hat{V}$ uniquely maximizes $\min_i N_i^{*\top} W_i$ while attaining
    the exact minimax bound $\hat{V}_i^\top m = \mu$ uniformly across all $i$, amplifying the
    signal-to-common-mode ratio by $\sqrt{\rho K/(1-\rho)}\bigl(1 + {\cal O}(K^{-1})\bigr)$ (${>}\,11\times$ in amplitude,
    ${>}\,135\times$ in variance at $K=12$) and canceling $\mu\gamma_t$ \emph{identically} in all
    pairwise dominance comparisons $(\hat{V}_i - \hat{V}_j)^\top X = (V_i - V_j)^\top X$ and
    stationary rupture tests.
\end{enumerate}

In summary, throughout this paper we focus on the un-centered Stiefel approach because
preserving strict monotonicity with respect to narrative alignment is our primary concern
when measuring narrative prevalence: a stronger manifestation of a narrative in a text
must never yield a lower prevalence score. We are mindful of the caveats associated with
its few remaining drawbacks---namely, the ${\cal O}(1/\sqrt{K})$ baseline offset in
absolute prevalence levels, the ${\cal O}(1/K)$ co-occurrence floor, and a mild
${\cal O}(1/K)$ cross-narrative demeaning shift---which diminish as the size $K$ of the
monitored collection grows and cancel out in relative dominance comparisons and
stationary rupture detection.

\section{Experiments and Validation on a Dataset}
\label{sec:application}

The objective of this empirical study is to validate the methodology introduced in the previous sections 
{on a labeled dataset} from two complementary perspectives. 
It is not obvious to find a dataset of text that has labels compatibles with narratives; we found the \emph{News Category Dataset} (described in detail in Section \ref{sec:application:data}).
What we need is a corpus of texts classified in theme corresponding to potential narratives so that we can check:
our orthogonalization procedure produce time series of prevalences that ``make sense'' (that is a qualitative procedure conducted in Section \ref{sec:application:qualitative}), 
by choosing narratives that are meant to be disjoint with respect to the labeled themes, we can check our orthonormalization procedure respect this \emph{disjunction} (typically orthogonalized narratives in the theme ``Entertainment'' does not align with orthogonalized narratives in the theme ``Politics'').

More precisely we will compare those two approaches: 
\begin{itemize}
    \item We use the embedding that directly encodes the text representations of the studied collection of narratives. We denote them as $(N_i)_i$ and we will sometimes refer to them as \og raw narratives \fg. 
    \item We use the vectors obtained after applying the procedure described in \ref{sec:ortho:stiefel}. We denote them as $(V_i)_i$ and we will sometimes refer to them as \og pure narratives \fg or \og orthogonalized narratives \fg. 
\end{itemize}

We want to check that: 
\begin{itemize}
    \item That orthogonalized narratives effectively remove the biases highlighted in ~\ref{sec:naive-failure};
    \item The orthogonalized representatives preserve the meaning;
    \item Tracking a document's similarity to an orthogonalized narrative is a more meaningful way to monitor the exposure to that specific narrative (compared to using the raw narratives)
\end{itemize}

\subsection{Dataset and Experimental Setup}
\label{sec:application:data}

\paragraph{Dataset.}
We use the \emph{News Category Dataset}~\autocite{misra2022news}, which contains approximately $210\,000$ headlines published by \emph{HuffPost} between 2012 and 2022. 
This dataset has been for instance used by \autocite{fewshot2024augmentation} for evaluating few-shot selection strategies in LLM-based text augmentation, 
\autocite{mediabias2024generated} to fine-tunes RoBERTa classifiers.

We perform descriptive statistics on this dataset to keep its main categories over the 42 available; it left us with 4 main categories: \emph{Politics} ($32\,739$ articles), \emph{Entertainment} ($16\,058$),
\emph{Travel} ($9\,887$), and \emph{Parenting} ($8\,677$ articles). These four chapters naturally match the narrative groups introduced below and provide labeled subsets that can be exploited for validation.
Since the archive becomes sparse after May 2018, we restrict the analysis to the 2012--2018 period, resulting in approximately $61\,235$ articles.

\paragraph{Temporal resampling.}
The raw dataset is highly unbalanced over time, particularly for the Politics category during the 2016 presidential campaign. To avoid interpreting variations in publication volume as changes in narrative prevalence, we perform a temporal resampling that enforces a constant weekly number of articles while preserving the relative proportions across categories.Table~\ref{tab:resampled} reports the resulting distribution. As shown in Table \ref{tab:resampled}: Entertainment and Politics are the ones stationary enough to test our temporal conditioning, and we keep Parenting and Travel for comparison.

\begin{table}[hbt!]
\begin{threeparttable}
\caption{Number of articles per category and year after resampling.}
\label{tab:resampled}
\begin{tabular}{lrrrr}
\toprule
\headrow Year & Entertainment & Parenting & Politics & Travel \\
\midrule
2012 &   792 & 4{,}087 &         0 & 4{,}221 \\
2013 & 1{,}154 & 4{,}710 &         0 & 3{,}786 \\
2014 & 2{,}445 & 1{,}403 & 3{,}684 & 2{,}079 \\
2015 & 3{,}614 &         0 & 5{,}351 &    705 \\
2016 & 2{,}478 &         0 & 6{,}885 &    311 \\
2017 & 2{,}194 &         0 & 7{,}273 &    178 \\
2018 & 1{,}375 &         0 & 2{,}450 &     60 \\
\bottomrule
\end{tabular}
\end{threeparttable}
\end{table}

\paragraph{Embedding.}
Both news headlines concatenated with article descriptions and narrative descriptions are embedded in the same $768$-dimensional semantic space using \emph{ModernBERT}~\autocite{warner2024modernbert}. Narrative descriptions were generated with the assistance of a large language model to ensure a homogeneous writing style, following emerging protocols for LLM-assisted narrative analysis~\autocite{schmidt2025identifying}.

\subsection{Narratives Tailored to the 2012--2018 Labeled Corpus}
\label{sec:application:narratives}

Because the \emph{News Category Dataset} covers the historical period 2012--2018 and is organized around HuffPost's editorial sections, the post-2022 policy narratives used for geometric illustration in Sections~\ref{sec:model} and~\ref{sec:ortho} (Appendix~\ref{app:narratives_illustration}, which reference generative AI language models and the 2022 Russia--Ukraine war) cannot be evaluated on this archive. Instead, to exploit the dataset's ground-truth category labels (\emph{Politics}, \emph{Entertainment}, \emph{Parenting}, and \emph{Travel}) as a counterfactual validation benchmark, we construct a dedicated \textbf{Historical HuffPost Validation Collection} of twelve narratives---three per labeled category---capturing the defining social, political, and cultural themes of the 2012--2018 period. The complete list of all twelve validation narratives is provided in Appendix~\ref{app:narratives} (Table~\ref{tab:narratives2}), and a representative selection is shown in Table~\ref{tab:narratives_short}.

\begin{table}[hbt!]
\begin{threeparttable}
\caption{Narratives used in the analysis (selection).}
\label{tab:narratives_short}
\begin{tabular}{llp{6cm}}
\toprule
\headrow Chapter & Narrative & Description \\
\midrule
Politics
  & Trump as Disruptor and Institutional Stress
  & Donald Trump rises to power by channeling distrust toward political elites,
    media institutions, and bureaucratic authority, presenting himself as a
    disruptive outsider challenging democratic norms. \\
\addlinespace
Entertainment
  & Entertainment as Political Resistance
  & Film, television, comedy, and celebrity culture become arenas of political
    contestation blending entertainment with explicit political messaging. \\
\addlinespace
Parenting
  & The Overwhelmed Modern Parent
  & Parenthood is portrayed as emotionally intense and structurally unsupported,
    with parents struggling to balance careers, childcare costs, and social
    expectations. \\
\addlinespace
Travel
  & Experiential and Transformational Travel
  & Travel is framed as a path to personal growth and cultural immersion through
    encounters with landscapes, heritage sites, and local communities. \\
\bottomrule
\end{tabular}
\end{threeparttable}
\end{table}

The $(N_i)_i$ raw narrative embedding are then orthogonalized using the Stiefel optimization procedure described in Section~\ref{sec:ortho}, producing the corresponding \textit{collection of pure narratives} $(V_i)_i$.

\subsection{First qualitative results}
\label{sec:application:qualitative}

We first perform a qualitative comparison between the prevalence estimators obtained using the raw narrative vectors and their orthogonalized representatives.

\begin{figure}[h!]
    \centering
    \includegraphics[width=0.7\textwidth]{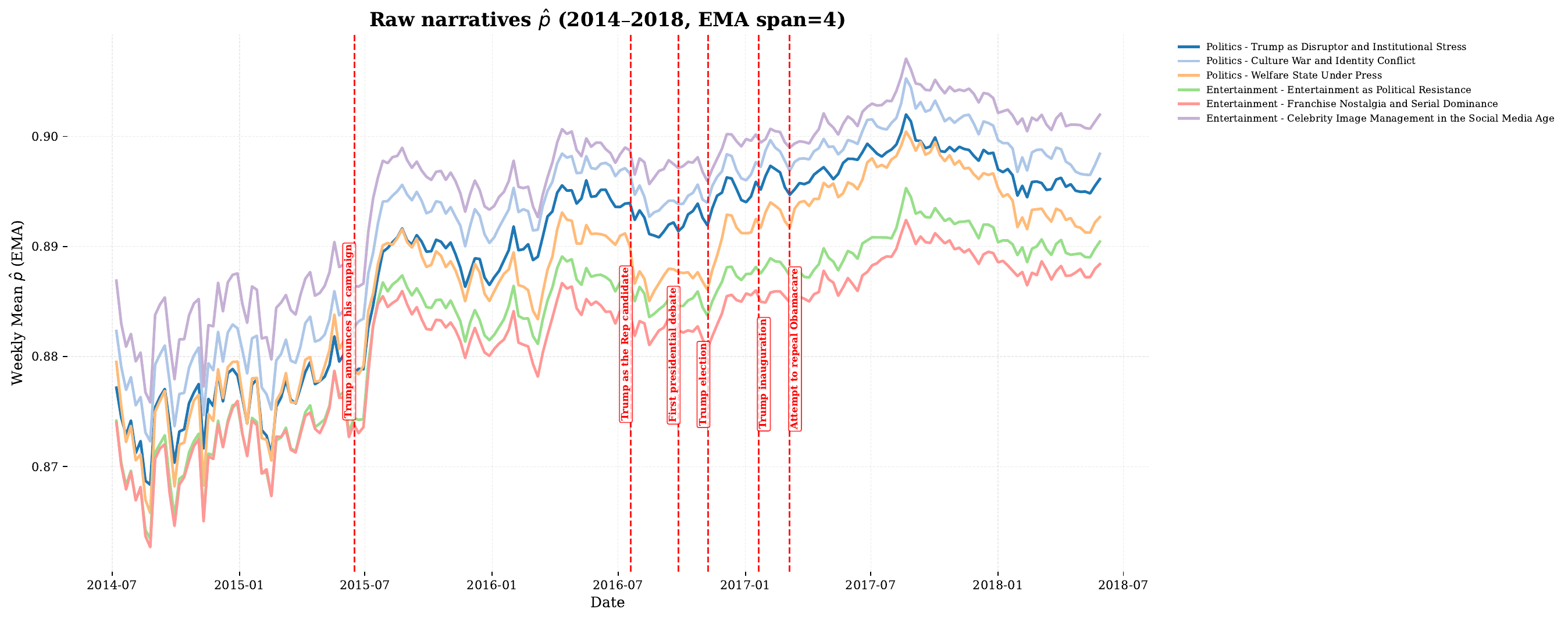}
    \vspace{0.3cm}

    \includegraphics[width=0.7\textwidth]{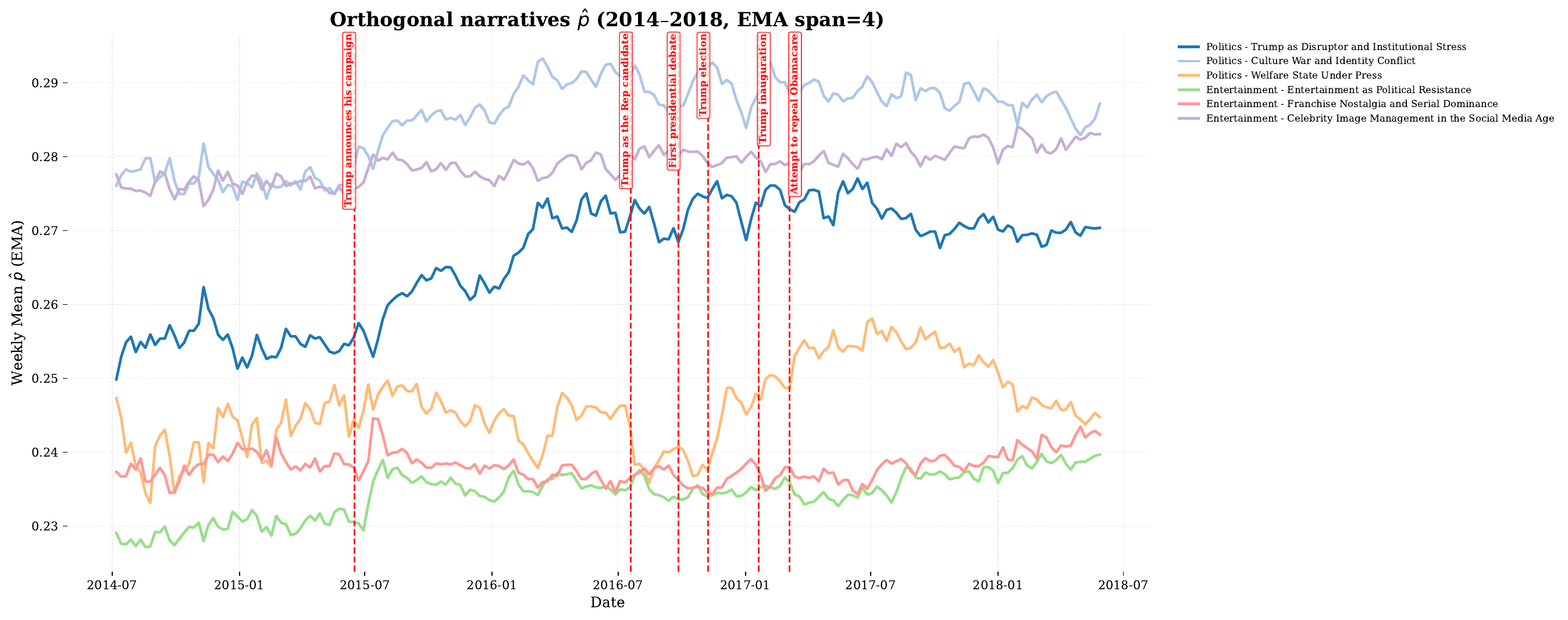}
    \caption{Time series of EMA weekly means of cosine similarities for raw and orthogonal narratives.}
    \label{fig:ema_comparison}
\end{figure}

To facilitate visual interpretation that can be done on Figure \ref{fig:ema_comparison}, we restrict this analysis to the Politics and Entertainment chapters over the 2014--2018 period, where both categories contain a sufficiently balanced number of observations. The displayed curves correspond to an exponential moving average of the estimator $\hat p_i(t)$ introduced in Section~\ref{sec:model}.

The contrast between the two approaches is immediate. Under the raw narrative vectors, the prevalence series evolve almost perfectly synchronously, illustrating the leakage phenomenon anticipated in Section~\ref{sec:model}: when the prevalence of an unrelated but semantically connected narrative rises, the naive prevalence indicator computed on raw narratives increases mechanically along with it. Pure narratives, by contrast, produce clearly differentiated temporal trajectories, with each of them now exhibiting its own distinctive dynamics.

The clearest visual illustration of this contrast occurs between January 2017 and January 2018, when a pronounced bump appears in the raw prevalence series of virtually every narrative. Under the purified representation, this bump is observed almost exclusively in the \narr{Welfare State Under Pressure} narrative, the only one for which it plausibly reflects genuine content rather than a shift shared indiscriminately across the corpus. 
\emph{This single episode captures the practical interest of orthogonalization: without it, an apparent surge in any given narrative could just as easily be an artifact of another, unrelated narrative's rise}. 
This pattern is consistent with the historical record, as the early months of the Trump administration were marked by sustained legislative pressure on the social safety, most notably through repeated attempts to repeal the Affordable Care Act (Obamacare).

The observed behavior is also consistent with the historical context used to design the narratives. The political narratives were deliberately chosen to capture the rise of Donald Trump's presidential campaign and its associated political themes. Accordingly, the prevalence of the narratives \narr{Trump as Disruptor and Institutional Stress} and \narr{Culture War and Identity Conflict} increases steadily throughout the campaign and reaches its highest levels around the 2016 election. This evolution is qualitatively consistent with the growing importance of these themes in the political debate.

Conversely, the Entertainment narratives were intentionally designed to represent long-term structural trends rather than isolated historical events. Their estimated prevalence therefore remains comparatively stable over time, exhibiting only moderate fluctuations. The resulting time series are thus coherent with the expected temporal behavior of the corresponding narratives.

\subsection{From Qualitative to Quantitative Virtues of Stiefel Orthonormalizaton}
\label{sec:application:quantitative}

We next evaluate the proposed methodology through three complementary quantitative experiments.

\subsubsection{Reduction of artificially high co-occurrence}
We first observe qualitatively that the prevalence curves within the political and Entertainment subsets of narratives evolve almost synchronously. We now turn to quantifying this phenomenon across the full set of narratives under study.

We use the metric introduced in Section~\ref{sec:model}:
\[
M_{ij}(t) := \mathbb{E}\bigl[(X^\top W_i)(X^\top W_j)\mid\mathcal{F}_t\bigr].
\]
As discussed there, for a collection of raw, non-orthogonalized narratives, two distinct effects artificially inflate this co-occurrence: narrative leakage and the correlation induced by the noise term. We therefore assess whether orthogonalization produces a sizable reduction in co-occurrence. Time is discretized weekly: for each pair $(i,j)$ we compute $M_{ij}(t)$ and average it over the entire study period.

\begin{table}[t]
\centering
\begin{tabular}{lcc}
\toprule
 & N (raw) & V (purified) \\
\midrule
mean & 0.809 & 0.071 \\
std & 0.007 & 0.008 \\
median & 0.808 & 0.070 \\
min & 0.795 & 0.057 \\
max & 0.825 & 0.091 \\
\bottomrule
\end{tabular}
\caption{Descriptive statistics of pairwise narrative co-movement $M_{ij}$, before (N) and after (V) orthogonalization.}
\label{tab:covariation_stats}
\end{table}

The empirical results strongly support this prediction: as reported in Table~\ref{tab:covariation_stats}, pairwise co-occurrence scores computed from the raw narrative vectors are, on average, approximately one order of magnitude larger than those obtained after orthogonalization.

\subsubsection{Semantic specificity}

The second validation exercise aims to verify that purified representatives preserve the semantic meaning of the original narratives.

The labeled structure of the \emph{News Category Dataset} naturally provides a counterfactual framework: since every narrative belongs to one of four chapters, we compare its average cosine similarity within its own chapter to its average similarity outside that chapter.

For the orthogonalized representatives, we define
\[
\Delta^{i,k}(V_{i,k})
=
\mathbb{E}
\!\left[
(X^{i})^\top V_{i,k}
\right]
-
\mathbb{E}
\!\left[
(X^{\neg i})^\top V_{i,k}
\right],
\]
where $X^{i}$ denotes articles belonging to chapter $i$, while $X^{\neg i}$ denotes all remaining articles.

The analogous quantity is computed for the original narrative vectors,
\[
\Delta^{i,k}(N_{i,k})
=
\mathbb{E}
\!\left[
(X^{i})^\top N_{i,k}
\right]
-
\mathbb{E}
\!\left[
(X^{\neg i})^\top N_{i,k}
\right].
\]

A meaningful narrative should be more strongly represented within its own chapter than outside it, implying
\[
\Delta^{i,k}(N_{i,k})>0,
\qquad
\Delta^{i,k}(V_{i,k})>0.
\]

More importantly, if orthogonalization successfully removes semantic information shared with other chapters while preserving each narrative's intended meaning, we expect
\[
\Delta^{i,k}(V_{i,k})
>
\Delta^{i,k}(N_{i,k}).
\]

\begin{table}[htbp]
\centering
\caption{Sensitivity analysis: 1000 bootstrap draws, 100\% of articles, 95\% CI for $\Delta^{i}(V_{i})$, $\Delta^{i}(N_{i})$, and $\Delta_{\text{gain}} = \Delta^i(V_i) - \Delta^i(N_i)$. Having all $\Delta^{i}(V_{i})\geq 0$ is compatible with our theory, whereas the only occurrences of unexpected $\Delta_{\text{gain}}$, i.e. negative values, are in \textcolor{red}{red}.}
\label{tab:sensitivity_analysis}
\small
\setlength{\tabcolsep}{4pt}
\renewcommand{\arraystretch}{1.15}
\resizebox{\textwidth}{!}{%
\begin{tabular}{llrrr}
\toprule
\textbf{Chapter} &
\textbf{Narrative} &
$\boldsymbol{\Delta^i(N_i)}$ &
$\boldsymbol{\Delta^i(V_i)}$ &
$\boldsymbol{\Delta_{\text{gain}}}$ \\
\midrule
Politics &
Trump as disruptor and institutional stress &
$0.0019\, [0.0015, 0.0023]$ &
$0.0377\, [0.0374, 0.0381]$ &
$0.0358\, [0.0353, 0.0363]$ \\

Politics &
Culture war and identity conflict &
$-0.0002\, [-0.0006, 0.0003]$ &
$0.0312\, [0.0309, 0.0316]$ &
$0.0314\, [0.0309, 0.0319]$ \\

Politics &
Welfare state under pressure &
$-0.0053\, [-0.0058, -0.0048]$ &
$0.0084\, [0.0078, 0.0090]$ &
$0.0137\, [0.0134, 0.0141]$ \\

Entertainment &
Entertainment as political resistance &
$-0.0145\, [-0.0151, -0.0139]$ &
$0.0062\, [0.0058, 0.0066]$ &
$0.0207\, [0.0203, 0.0210]$ \\

Entertainment &
Franchise nostalgia and serial dominance &
$-0.0107\, [-0.0112, -0.0101]$ &
$0.0207\, [0.0202, 0.0211]$ &
$0.0313\, [0.0309, 0.0318]$ \\

Entertainment &
Celebrity image management in the social media... &
$-0.0153\, [-0.0159, -0.0148]$ &
$0.0054\, [0.0050, 0.0058]$ &
$0.0207\, [0.0202, 0.0212]$ \\

Parenting &
The overwhelmed modern parent &
$0.0253\, [0.0249, 0.0256]$ &
$0.0118\, [0.0114, 0.0123]$ &
\color{red}$-0.0134\, [-0.0139, -0.0129]$ \\

Parenting &
Intensive and expert guided parenting &
$0.0287\, [0.0283, 0.0291]$ &
$0.0302\, [0.0297, 0.0308]$ &
$0.0015\, [0.0010, 0.0020]$ \\

Parenting &
Raising children in a politicized society &
$0.0244\, [0.0240, 0.0248]$ &
$0.0174\, [0.0171, 0.0178]$ &
\color{red}$-0.0070\, [-0.0073, -0.0066]$ \\

Travel &
Experiential and transformational travel &
$0.0251\, [0.0246, 0.0257]$ &
$0.0298\, [0.0293, 0.0304]$ &
$0.0047\, [0.0042, 0.0052]$ \\

Travel &
Travel optimization and consumer strategy &
$0.0279\, [0.0274, 0.0285]$ &
$0.0415\, [0.0410, 0.0420]$ &
$0.0135\, [0.0130, 0.0141]$ \\

Travel &
Aesthetic and instagrammable destinations &
$0.0297\, [0.0293, 0.0303]$ &
$0.0452\, [0.0446, 0.0458]$ &
$0.0155\, [0.0150, 0.0160]$ \\
\bottomrule
\end{tabular}%
}
\end{table}

The results in Table \ref{tab:sensitivity_analysis} are satisfactory: every chapter exhibits $\Delta^i(V_i) > 0$ 
meaning that all orthogonalized representatives are more strongly prevalent within their own chapter than outside it.
The \emph{Entertainment} chapter and two narratives of \emph{Politics} chapter have $\Delta^i(N_i) < 0$. Regarding $\Delta_{\text{gain}}$, it is positive for three chapters out of four, meaning that in the majority of cases, the orthogonalization process successfully preserves the narrative's meaning while removing the semantic component shared with other chapters.

\subsubsection{Narrative specificity of detected ruptures}
\label{sec:application:quantitative:rupture}

We now quantify the qualitative observation made above: ruptures detected on raw narrative representations tend to be shared across many, otherwise unrelated, narratives, whereas ruptures detected on purified representations are largely specific to a single narrative.

\begin{figure}[h!]
    \centering
    \includegraphics[width=0.7\textwidth]{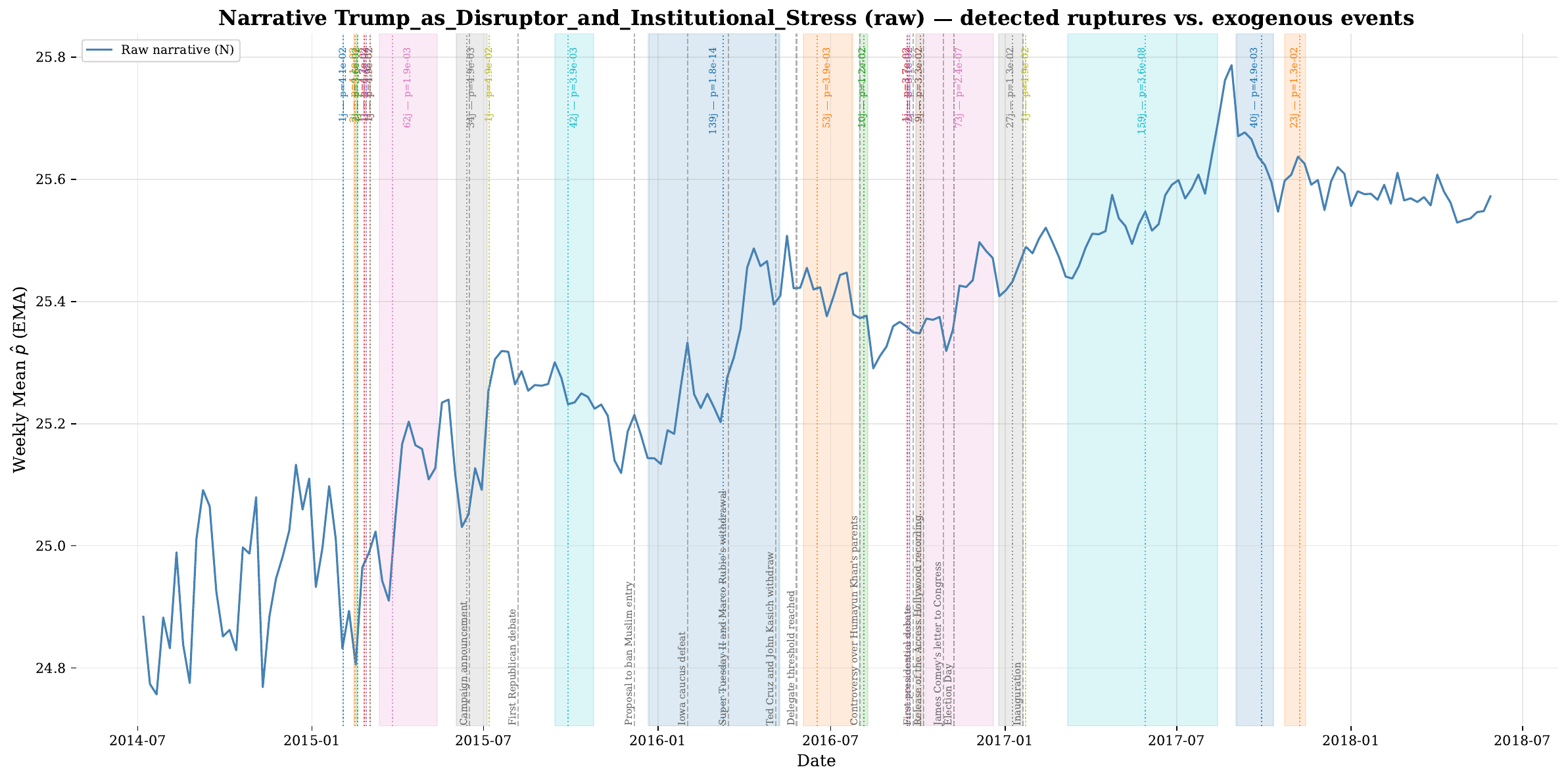}
    \vspace{0.3cm}

    \includegraphics[width=0.7\textwidth]{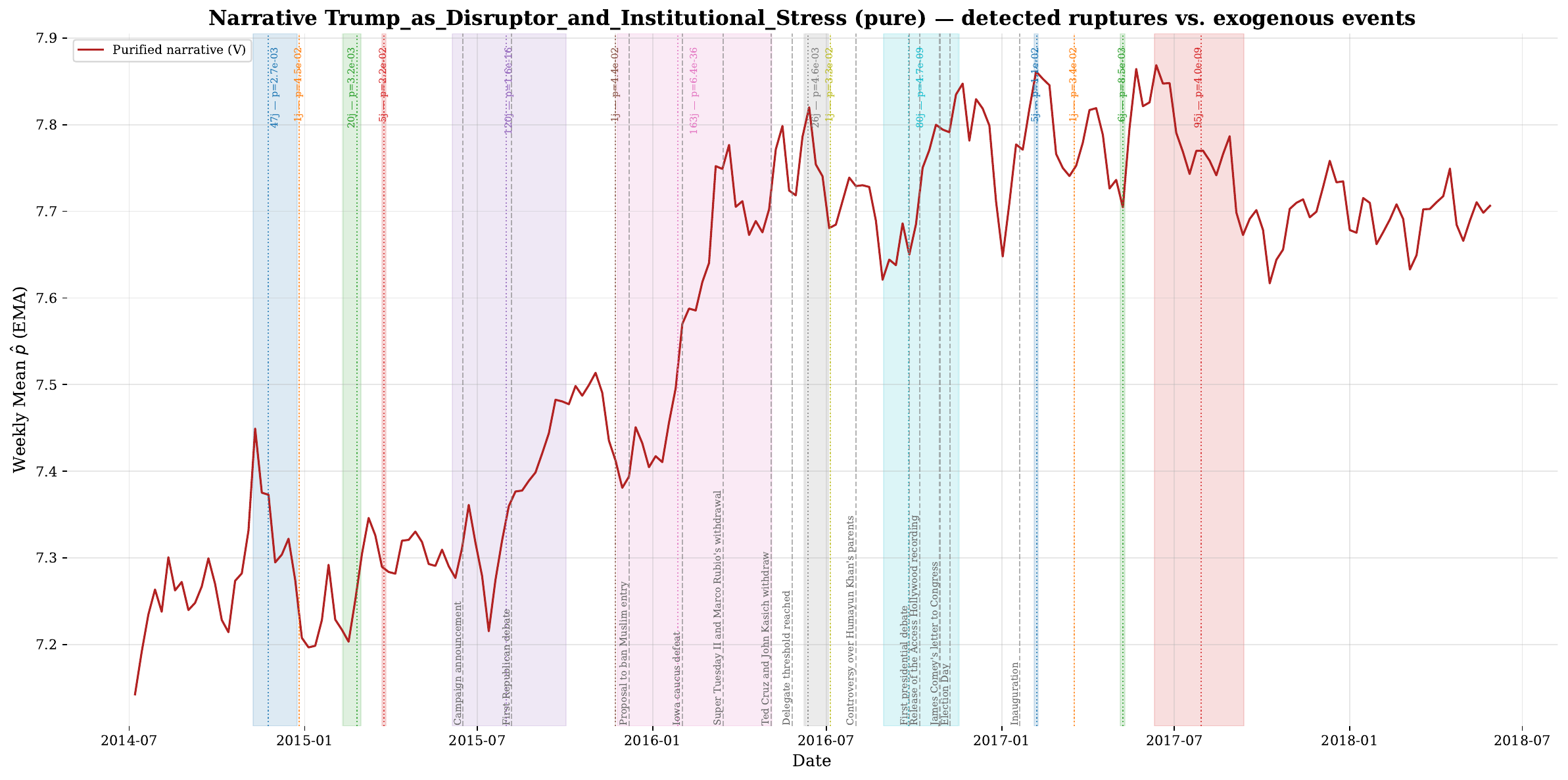}
    \caption{Time series of EMA weekly means of cosine similarities for raw and orthogonal Trump narrative, with detected ruptures as well as a set of exogenous events related to Trump's campaign.} 
    \label{fig:rupture_comparison}
\end{figure}

We define a rupture by the rejection by a two-sided Welch $t$-test of the null hypothesis:\footnote{Because this test is repeated over hundreds of candidate dates, the resulting $p$-values are corrected for multiple testing using the Benjamini-Hochberg FDR procedure; consecutive significant dates are then grouped into a single detected rupture} ``\emph{the prevalence of the narrative $M$ in News before $t$ is statistically not different from the prevalence of the narrative $M$ in News after $t$}''.
We define $\mathcal{B}_t$ and $\mathcal{A}_t$ a windows of 90 days respectively before and after $t$ and we test, on the raw narratives $N$ and on the pure narratives $V$ the hypothesis at each date $t$:
\[
H_0(t, M) : \mathbb{E}\bigl[X^\top M \mid X \in \mathcal{B}_t\bigr] = \mathbb{E}\bigl[X^\top M \mid X \in \mathcal{A}_t\bigr].
\]
It means that we obtain one time series $H_0(t, N_i)$ for each raw narrative $N_i$ (upper panel of Figure \ref{fig:rupture_comparison}) and one time series $H_0(t, V_i)$ for each corresponding pure narrative $V_i$ (lower panel of Figure \ref{fig:rupture_comparison}).
This procedure is applied separately to the raw ($N$) and purified ($V$) representations of every narrative in our collection.

Complete rupture-detection plots for all twelve validation narratives are reported in Appendix~\ref{app:ruptures} (Figure~\ref{fig:ruptures}). In Figure~\ref{fig:rupture_comparison}, we display the detailed time series for the narrative \narr{Trump as Disruptor and Institutional Stress}.

Three observations, all compatible with our model, comforting our methodology, stand out:
\begin{itemize}
    \item First, the number of detected ruptures is systematically higher under $N$ than under $V$: averaged across narratives, $27.1$ ruptures are detected under $N$, against $16.5$ under $V$, and this ordering holds narrative by narrative without exception.
    
    \item Second, pairwise overlap between the sets of rupture dates detected across different narratives is markedly higher under $N$ than under $V$, as measured by the Jaccard index\footnote{Jaccard index: $J(A,B) = \dfrac{|A \cap B|}{|A \cup B|}$.}: the average pairwise Jaccard index is $0.74$ under $N$, compared to only $0.25$ under $V$. Taken together, these two findings confirm that raw representations tend to co-detect the same, corpus-wide ruptures, consistent with the leakage phenomenon described in Section~\ref{sec:naive-failure}, whereas purified representations isolate ruptures that are substantially more specific to each individual narrative.
    
    \item Third, this is directly illustrated by the July 2017 bump discussed in our qualitative analysis: under $N$, it is detected as a significant rupture for every narrative in the collection, while under $V$ it is detected almost exclusively for the \narr{Welfare State Under Pressure} narrative, the only one for which it plausibly reflects genuine, narrative-specific content (as the period corresponds to Trump's first semester as President with notably repeated attempt to repeal the Obamacare) rather than a shift shared indiscriminately across the corpus.
\end{itemize}
\bigskip

\section{Conclusion}
\label{sec:conclusion}

We have introduced a rigorous algebraic methodology for monitoring the relative
prevalence of competing narratives in a time-evolving text corpus.

It allows to estimate the relative prevalence of narratives directly in the space of embedding, that is far cheaper and more reliable than prompting frontier models.

The key insight is that standard cosine similarity 
{between embedding of raw narrative (i.e. arbitrary representative of a narrative using a sentence in natural language)} suffers from a structural correlation bias induced
by shared semantic content, which we characterize formally through a linear
generative model.
It allows us to propose a method to orthogonalize arbitrary collection of narratives to turn them into what we call \emph{a collection of pure narratives}.
Our remedy is a global orthogonalization on the Stiefel manifold. 
It eliminates
the shared component while preserving each narrative's idiosyncratic direction,
yielding an unbiased, non-parametric, and interpretable prevalence estimator.
Validation on a large labeled news corpus confirms that the estimator
faithfully recovers known historical dynamics, whereas the naive approach fails.

Using a labelled dataset (the \emph{News Category Dataset}), we show qualitatively and quantitatively that our method is sound and allow to draw conclusions that otherwise would have been hidden into artifacts of natural language syntax. 

\paragraph{Perspectives.}
The natural extension of this work is the application of the estimator to
large-scale 
{economic} corpora, opening the way to a direct quantitative
model of narrative competition and its influence on the formation of
economic expectations~\autocite{azqueta2026diverging}, that is a question at the intersection of computational linguistics
and price formation.

\paragraph{Acknowledgments.}
Authors thank Emile Servan-Schreiber for numerous discussions on the intersection of narratives and prediction markets, especially in the context of the use of language models to synthesize artificial competition between narratives.

\paragraph{Competing Interests.}
None.

\paragraph{Data Availability Statement.}
The News Category Dataset is publicly
available~\autocite{misra2022news}. Replication code can be found at
{\tt https://github.com/ca-lehalle/narrative-geometry}.

\paragraph{Ethical Standards.}
The research meets all ethical guidelines, including adherence to the legal requirements of the study country.

\paragraph{Author Contributions.}
The initial work on orthogonalization in the space of embedding of narrative started as a student project of 3rd year of Ecole Polytechnique by Juliette Anglade and Mario Massy under the direction of Pr. Charles-Albert Lehalle. It later evolved in a joined research effort, reported into this paper. The contributions of the three authors is thus equal.

\printbibliography

\appendix

\section{Contemporary Policy Narratives (Geometric Illustration in Sections~\ref{sec:model}--\ref{sec:ortho})}
\label{app:narratives_illustration}

\begin{longtable}{p{2cm}p{4cm}p{7cm}}
\caption{The twelve contemporary macro-policy narratives (four chapters $\times$ three narratives) used to illustrate embedding anisotropy and Stiefel vs.\ Gram--Schmidt orthogonalization in Sections~\ref{sec:model}, \ref{sec:method}, and~\ref{sec:ortho}.}
\label{tab:narratives}\\
\toprule
\textbf{Chapter} & \textbf{Narrative} & \textbf{Description} \\
\midrule
\endfirsthead

\multicolumn{3}{c}{\tablename\ \thetable{} -- \textit{Continued from previous page}} \\
\toprule
\textbf{Chapter} & \textbf{Narrative} & \textbf{Description} \\
\midrule
\endhead

\midrule
\multicolumn{3}{r}{\textit{Continued on next page}} \\
\endfoot

\bottomrule
\endlastfoot

Climate
  & Extreme Heat and Emergency Policy Response
  & Record-breaking temperatures across southern Europe this summer triggered
    unprecedented heatwaves, forcing governments to activate emergency climate plans
    and accelerate the transition toward renewable energy sources in order to
    meaningfully reduce carbon emissions and limit further global warming. \\
Climate
  & International Climate Commitments Under Pressure
  & As climate-related disasters multiply worldwide, the United Nations urgently calls
    for stronger adherence to the Paris Agreement, emphasizing that industrialized
    nations must deliver substantial and verifiable reductions in greenhouse gas
    emissions before irreversible climate tipping points are reached. \\
Climate
  & Rising Seas and Coastal Vulnerability
  & The accelerating melting of Arctic glaciers, driven by sustained global warming,
    is causing a gradual but measurable rise in sea levels that threatens tens of
    millions of people living in low-lying coastal zones, according to a major
    scientific report released this week. \\
\addlinespace
Artificial Intelligence
  & AI Disruption of Labor Markets
  & The rapid rise of generative artificial intelligence models is profoundly reshaping
    global labor markets, prompting technology companies to invest billions of dollars
    in autonomous systems capable of gradually replacing skilled workers across a wide
    range of industries and professional sectors. \\
Artificial Intelligence
  & Big Tech Race and Regulatory Pushback
  & Silicon Valley's leading technology firms are intensifying their competition in
    artificial intelligence by unveiling increasingly powerful language models, while
    European regulators work to constrain these developments through strict legislative
    frameworks designed to protect citizens' fundamental rights and digital freedoms. \\
Artificial Intelligence
  & AI Ethics, Privacy, and Public Sector Governance
  & Recent breakthroughs in artificial intelligence are raising critical ethical
    concerns around personal data privacy and cybersecurity, prompting several national
    governments to establish dedicated regulatory frameworks governing the deployment
    of these technologies within public institutions and essential services. \\
\addlinespace
Geopolitics
  & NATO Eastern Flank and the Russia--Ukraine War
  & The ongoing military conflict between Russia and Ukraine continues to destabilize
    the European geopolitical order, compelling NATO allies to reinforce their eastern
    flank with additional troops and to significantly increase national defense budgets
    in response to sustained and escalating Russian aggression. \\
Geopolitics
  & Middle East Conflict and Humanitarian Crisis
  & The armed conflict in the Middle East has triggered a humanitarian catastrophe of
    alarming proportions, prompting urgent international calls for an immediate
    ceasefire and the rapid deployment of emergency aid to civilian populations trapped
    in active combat zones across the region. \\
Geopolitics
  & US--China Rivalry and Indo-Pacific Tensions
  & Growing geopolitical rivalry between Washington and Beijing over the status of
    Taiwan is seriously threatening stability across the Indo-Pacific region, with both
    superpowers escalating military demonstrations and economic sanctions in what
    increasingly resembles a renewed cold war confrontation. \\
\addlinespace
Economics
  & Inflation, Interest Rates, and Household Strain
  & Persistent inflation across the world's major economies is forcing central banks
    to maintain elevated interest rates, placing significant pressure on economic
    growth and disproportionately burdening lower-income households struggling to
    absorb the rising cost of everyday goods and essential services. \\
Economics
  & Financial Market Volatility and Fed Signaling
  & Global financial markets experienced sharp volatility this week as investors
    reacted to the latest monetary policy signals from the United States Federal
    Reserve, triggering significant fluctuations across major stock indices and a
    notable depreciation of several emerging market currencies. \\
Economics
  & IMF Forecasts and Global Growth Slowdown
  & Global economic growth is decelerating at a concerning pace according to the
    latest International Monetary Fund projections, which identify trade tensions,
    rising public debt levels, and slowing emerging market performance as the primary
    drivers behind deteriorating worldwide economic prospects. \\
\end{longtable}

\section{Historical HuffPost Narratives (Empirical Validation in Section~\ref{sec:application})}
\label{app:narratives}

\begin{longtable}{p{2cm}p{4cm}p{7cm}}
\caption{The twelve historical narratives (four HuffPost categories $\times$ three narratives) tailored to the 2012--2018 \emph{News Category Dataset} for the qualitative and quantitative experiments in Section~\ref{sec:application} and Appendix~\ref{app:ruptures}.}
\label{tab:narratives2}\\
\toprule
\textbf{Chapter} & \textbf{Narrative} & \textbf{Description} \\
\midrule
\endfirsthead

\multicolumn{3}{c}{\tablename\ \thetable{} -- \textit{Continued from previous page}} \\
\toprule
\textbf{Chapter} & \textbf{Narrative} & \textbf{Description} \\
\midrule
\endhead

\midrule
\multicolumn{3}{r}{\textit{Continued on next page}} \\
\endfoot

\bottomrule
\endlastfoot

Politics
  & Trump as Disruptor and Institutional Stress
  & Donald Trump rises to power by channeling distrust toward political elites,
    media institutions, and bureaucratic authority, presenting himself as a disruptive
    outsider challenging democratic norms and reframing institutional resistance as
    partisan sabotage. \\
Politics
  & Culture War and Identity Conflict
  & American politics becomes increasingly structured around culture wars in which
    abortion, LGBTQ rights, immigration, race, religion, and gun control become
    symbolic battlegrounds mobilizing voters through moralized political conflict. \\
Politics
  & Welfare State Under Pressure
  & Political debates around health care reform, social assistance, and regulatory
    protections reflect a broader struggle over the future of the welfare state amid
    economic insecurity and ideological polarization. \\
\addlinespace
Entertainment
  & Entertainment as Political Resistance
  & Film, television, comedy, and celebrity culture become arenas of political
    contestation where cultural production increasingly blends entertainment with
    explicit political messaging. \\
Entertainment
  & Franchise Nostalgia and Serial Dominance
  & The entertainment industry relies heavily on sequels, reboots, and cinematic
    universes, capitalizing on nostalgia and established IP in an increasingly
    competitive streaming environment. \\
Entertainment
  & Celebrity Image Management in the Social Media Age
  & Celebrities navigate a hyper-visible media ecosystem where personal relationships,
    identity declarations, and scandals shape public image and reputation management. \\
\addlinespace
Parenting
  & The Overwhelmed Modern Parent
  & Parenthood is portrayed as emotionally intense and structurally unsupported,
    with parents struggling to balance careers, childcare costs, and social expectations. \\
Parenting
  & Intensive and Expert-Guided Parenting
  & Parenting becomes increasingly expert-driven, with reliance on psychological
    advice, scientific studies, and educational technologies to optimize children's
    development. \\
Parenting
  & Raising Children in a Politicized Society
  & Parents confront a social environment shaped by political polarization, gun
    violence debates, and shifting gender norms that influence children's values and
    civic identity. \\
\addlinespace
Travel
  & Experiential and Transformational Travel
  & Travel is framed as a path to personal growth and cultural immersion through
    encounters with landscapes, heritage sites, and local communities. \\
Travel
  & Travel Optimization and Consumer Strategy
  & Travel is increasingly treated as a strategic consumer activity involving price
    optimization, loyalty programs, and efficiency strategies. \\
Travel
  & Aesthetic and Instagrammable Destinations
  & Destinations gain visibility through their visual appeal and shareability on social
    media platforms, emphasizing photogenic landscapes and unique experiences. \\
\bottomrule
\end{longtable}

\section{Proofs}
\label{app:proofs}

Throughout, $\mathcal V=\mathrm{Vect}(V_1,\ldots,V_K)$, $\Pi_{\mathcal V^\perp}$ denotes the orthogonal
projector onto $\mathcal V^\perp$ (of dimension $d-K\ge 1$). With the convention of
Section~\ref{sec:model:prevalence}, $p_i(t)=\beta\,p_{i,t}$ and $p_{ij}(t)=\beta^2\,p_{ij,t}$, where
$p_{i,t}=\mathbb E[\alpha_i\mid\mathcal F_t]$ and $p_{ij,t}=\mathbb E[\alpha_i\alpha_j\mid\mathcal F_t]$
(so $p_{ii,t}=p_{i,t}$ since $\alpha_i^2=\alpha_i$).

\subsection{Preliminary facts}

\begin{enumerate}
  \item[(F1)] \emph{Unit norm and exact projections.} Since $\varepsilon\in\mathcal V^\perp$ and $(V_i)$ is orthonormal,
  \[
    ||X||^2=\beta^2\sum_k\alpha_k+||\varepsilon||^2=\beta^2\sum_k\alpha_k+\Bigl(1-\beta^2\sum_k\alpha_k\Bigr)=1,
  \]
  because $||u||=1$. Moreover $V_i^\top X=\beta\sum_k\alpha_kV_i^\top V_k+V_i^\top\varepsilon=\beta\,\alpha_i$.

  \item[(F2)] \emph{Centered residual.} $\sqrt{1-\beta^2 \sum_k \alpha_k}$ is independent of $u$ and $u$ is independent of $\mathcal F_t$, hence
  \[
    \mathbb E[\varepsilon\mid\mathcal F_t]=\Esp \left [\sqrt{1-\beta^2 \sum_k \alpha_k} \mid \calF_t \right]\,\mathbb E[u]=0,
  \]
  because the uniform law on a sphere is invariant under $u\mapsto-u$.

  \item[(F3)] \emph{Second moment of the residual.} Let $m:=d-K$ and let $(e_1,\ldots,e_m)$ be an
  orthonormal basis of $\mathcal V^\perp$; write $u=\sum_{a=1}^mu_ae_a$ with $u_a=e_a^\top u$, so that
  $\mathbb E[uu^\top]=\sum_{a,b}\mathbb E[u_au_b]\,e_ae_b^\top$. The uniform law on the unit sphere is invariant under
  every isometry of $\mathcal V^\perp$. Flipping the sign of the coordinate $u_a$ is such an isometry, and it
  changes $u_au_b$ into $-u_au_b$ for $b\neq a$; hence $\mathbb E[u_au_b]=0$ for $a\neq b$. Swapping two
  coordinates is also an isometry, so $\mathbb E[u_a^2]$ does not depend on $a$; since
  $\sum_au_a^2=||u||^2=1$, we get $\mathbb E[u_a^2]=1/m$, and
  $\mathbb E[uu^\top]=\frac1m\sum_ae_ae_a^\top=\Pi_{\mathcal V^\perp}/(d-K)$. Therefore, as $u$ is independent of $\alpha_k$ and $\calF_t$
  \[
    \mathbb E[\varepsilon\varepsilon^\top\mid\mathcal F_t]= \Esp \left [1-\beta^2 \sum_k \alpha_k \mid \calF_t \right]\,\frac{\Pi_{\mathcal V^\perp}}{d-K} =\sigma_t^2\,\Pi_{\mathcal V^\perp},
  \]
  with
  \[
    \sigma_t^2:=\frac{1-\beta^2\sum_{k}p_{k,t}}{d-K},
  \]
\end{enumerate}

\subsection{Proof of Proposition~\ref{prop:prevalence-projection}}

By (F1), $V_i^\top X=\beta\,\alpha_i$. Taking $\mathbb E[\cdot\mid\mathcal F_t]$ gives
$\mathbb E[V_i^\top X\mid\mathcal F_t]=\beta\,p_{i,t}=p_i(t)$. \qed

\subsection{Proof of Proposition~\ref{prop:cooccurrence-projection}}

By (F1), $(V_i^\top X)(V_j^\top X)=\beta^2\alpha_i\alpha_j$ exactly. Taking $\mathbb E[\cdot\mid\mathcal F_t]$ gives
$\beta^2p_{ij,t}=p_{ij}(t)$. \qed

\subsection{Proof of Proposition~\ref{prop:naive-prevalence}}

Let $N_i=\sum_kc_{ik}V_k+N_i^\perp$ as in~\eqref{eq:candidate-decomp}. Since $N_i^\perp\in\mathcal V^\perp$ and $V_k^\top V_l=\delta_{kl}$,
\[
  N_i^\top X=\beta\sum_kc_{ik}\alpha_k+(N_i^\perp)^\top\varepsilon .
\]
Taking $\mathbb E[\cdot\mid\mathcal F_t]$ and using (F2) for the second term,
\[
  \pi_{N_i}(t)=\beta\sum_kc_{ik}\,p_{k,t}=\sum_kc_{ik}\,p_k(t)=c_{ii}\,p_i(t)+\sum_{k\neq i}c_{ik}\,p_k(t). \qquad\qed
\]

\subsection{Proof of Proposition~\ref{prop:naive-cooccurrence}}

Let $W_i=\sum_kc_{ik}V_k+\eta_i$ and $W_j=\sum_lc_{jl}V_l+\eta_j$, with $\eta_i,\eta_j\in\mathcal V^\perp$. Then
\[
  X^\top W_i=\beta\sum_kc_{ik}\alpha_k+\eta_i^\top\varepsilon,
  \qquad
  X^\top W_j=\beta\sum_lc_{jl}\alpha_l+\eta_j^\top\varepsilon,
\]
and the product expands into
\[
  (X^\top W_i)(X^\top W_j)=\beta^2\sum_{k,l}c_{ik}c_{jl}\,\alpha_k\alpha_l
  +\beta\sum_kc_{ik}\alpha_k\,\eta_j^\top\varepsilon
  +\beta\sum_lc_{jl}\alpha_l\,\eta_i^\top\varepsilon
  +\eta_i^\top\varepsilon\varepsilon^\top\eta_j .
\]
We take $\mathbb E[\cdot\mid\mathcal F_t]$ term by term.
\begin{itemize}
  \item First term: $\beta^2\sum_{k,l}c_{ik}c_{jl}\,p_{kl,t}=\sum_{k,l}c_{ik}c_{jl}\,p_{kl}(t)$.
  \item Cross terms: by (F2),
  $\mathbb E[\alpha_k\,\eta_j^\top\varepsilon\mid\mathcal F_t] = 0$, and likewise for the other cross term.
  \item Last term: by (F3) and $\Pi_{\mathcal V^\perp}\eta_j=\eta_j$,
  $\mathbb E[\eta_i^\top\varepsilon\varepsilon^\top\eta_j\mid\mathcal F_t]=\sigma_t^2\,\eta_i^\top\eta_j$.
\end{itemize}
Hence, exactly,
\begin{equation}
  \label{eq:naive-cooccurrence-exact}
  M_{ij}(t)=\sum_{k,l=1}^Kc_{ik}\,c_{jl}\,p_{kl}(t)+\sigma_t^2\,\eta_i^\top\eta_j .
\end{equation}
For $i\neq j$, isolating the pairs $(k,l)=(i,j)$ and $(j,i)$ (recall $p_{ji}=p_{ij}$),
\[
  M_{ij}(t)=\bigl(c_{ii}c_{jj}+c_{ij}c_{ji}\bigr)\,p_{ij}(t)
  +\underbrace{\sum_{(k,l)\notin\{(i,j),(j,i)\}}c_{ik}c_{jl}\,p_{kl}(t)}_{\text{narrative leakage}}
  +\underbrace{\sigma_t^2\,\eta_i^\top\eta_j}_{\text{noise term}},
\]
where the leakage sum includes the diagonal terms $k=l$, for which $p_{kk}(t)=\beta^2 p_{k,t}=\beta\,p_k(t)$ (since $\alpha_k^2=\alpha_k$). For pure candidates, $c_{ik}=\delta_{ik}$, $c_{jl}=\delta_{jl}$ and $\eta_i=\eta_j=0$, so $M_{ij}(t)=p_{ij}(t)$. \qed

\subsection{Unbiasedness of the empirical estimators}

Fix a period $t$, with $\mathcal C_t=\{n:\tau(\omega_n)=t\}$ and $n_t=|\mathcal C_t|$. Since $\tau$ is a stopping time, $\{\tau(\omega_n)=t\}\in\mathcal F_t$, so $\mathcal C_t$ and $n_t$ are $\mathcal F_t$-measurable. Implicitly, we make the following sampling assumption: for every $n$, on the event $\{\tau(\omega_n)=t\}$, the occurrence $\omega_n$ has, conditionally on $\mathcal F_t$, the law of the generic occurrence described in Section~\ref{sec:model:generative}. Estimators are considered on the event $\{n_t\ge1\}$.

\begin{proof}
Since $n_t$ and $\mathcal C_t$ are $\mathcal F_t$-measurable, they can be taken out of the conditional expectation, and by linearity
\[
  \mathbb E\bigl[\hat p_i(t)\mid\mathcal F_t\bigr]
  =\frac1{n_t}\sum_{n\in\mathcal C_t}\mathbb E\bigl[V_i^\top X_n\mid\mathcal F_t\bigr]
  =\frac1{n_t}\sum_{n\in\mathcal C_t}p_i(t)=p_i(t),
\]
where each term equals $p_i(t)$ by Proposition~\ref{prop:prevalence-projection} applied to $\omega_n$ (sampling assumption). The same argument with Proposition~\ref{prop:cooccurrence-projection} gives
$\mathbb E[\hat p_{ij}(t)\mid\mathcal F_t]=\frac1{n_t}\sum_{n\in\mathcal C_t}p_{ij}(t)=p_{ij}(t)$, and by the tower property both estimators are unbiased.
\end{proof}

\begin{remark}
The same computation shows that $\hat M_{ij}(t)=\frac1{n_t}\sum_{n\in\mathcal C_t}(X_n^\top W_i)(X_n^\top W_j)$ satisfies $\mathbb E[\hat M_{ij}(t)\mid\mathcal F_t]=M_{ij}(t)$, with $M_{ij}(t)$ given by~\eqref{eq:naive-cooccurrence-exact}.
\end{remark}

\section{Rupture Detection for All Validation Narratives}
\label{app:ruptures}

We display here the results of the experiment described in
Section~\ref{sec:application:quantitative:rupture}. For each narrative, we
display the time series of $\hat p(t)$ both for the raw ($N$) and the pure
representative of the narrative ($V$). We rescale the time series of the
projection on the pure narrative to observe both on the same $y$-axis. We
display as blue (resp.\ red) rectangles the detected ruptures when performing
the test described in Section~\ref{sec:application:quantitative:rupture}.

The gray, dotted vertical lines correspond to the following list of events that occurred during Trump's campaign. They were chosen exogenously and independently. They come from \href{https://www.cjr.org/special_report/trump-coverage-election-clinton.php}{Columbia Journalism Review}\footnote{\url{https://www.cjr.org/special_report/trump-coverage-election-clinton.php}}. 

\begin{table}[H]
  \centering
  \caption{Key events of the 2015--2017 Trump campaign.}
  \label{tab:trump-events}
  \begin{tabular}{@{}ll@{}}
    \toprule
    \textbf{Date} & \textbf{Event} \\
    \midrule
    2015-06-16 & Campaign announcement \\
    2015-08-06 & First Republican debate \\
    2015-12-07 & Proposal to ban Muslim entry \\
    2016-02-01 & Iowa caucus defeat \\
    2016-03-15 & Super Tuesday II and Marco Rubio's withdrawal \\
    2016-05-04 & Ted Cruz and John Kasich withdraw \\
    2016-05-26 & Delegate threshold reached \\
    2016-08-01 & Controversy over Humayun Khan's parents \\
    2016-09-26 & First presidential debate \\
    2016-10-07 & Release of the Access Hollywood recording \\
    2016-10-28 & James Comey's letter to Congress \\
    2016-11-08 & Election Day \\
    2017-01-20 & Inauguration \\
    \bottomrule
  \end{tabular}
\end{table}

\begin{figure}[!htbp]
  \centering
  \includegraphics[width=0.6\textwidth]{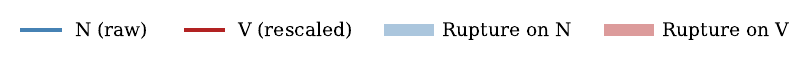}\par\vspace{0.6em}
  \begin{subfigure}[t]{0.49\textwidth}
    \centering
    \includegraphics[width=\linewidth]{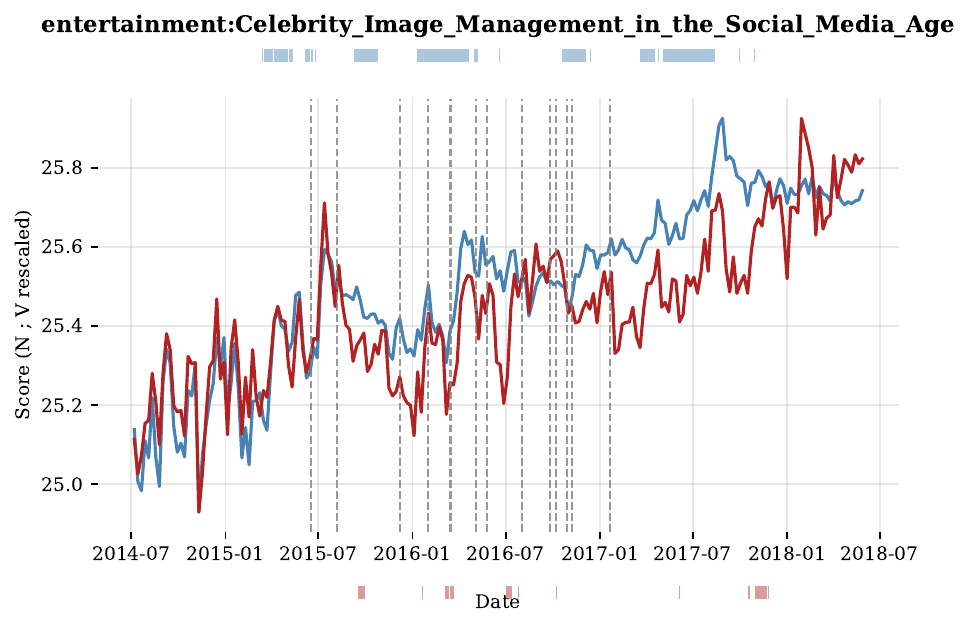}
    \caption{Narrative 1}
  \end{subfigure}
  \hfill
  \begin{subfigure}[t]{0.49\textwidth}
    \centering
    \includegraphics[width=\linewidth]{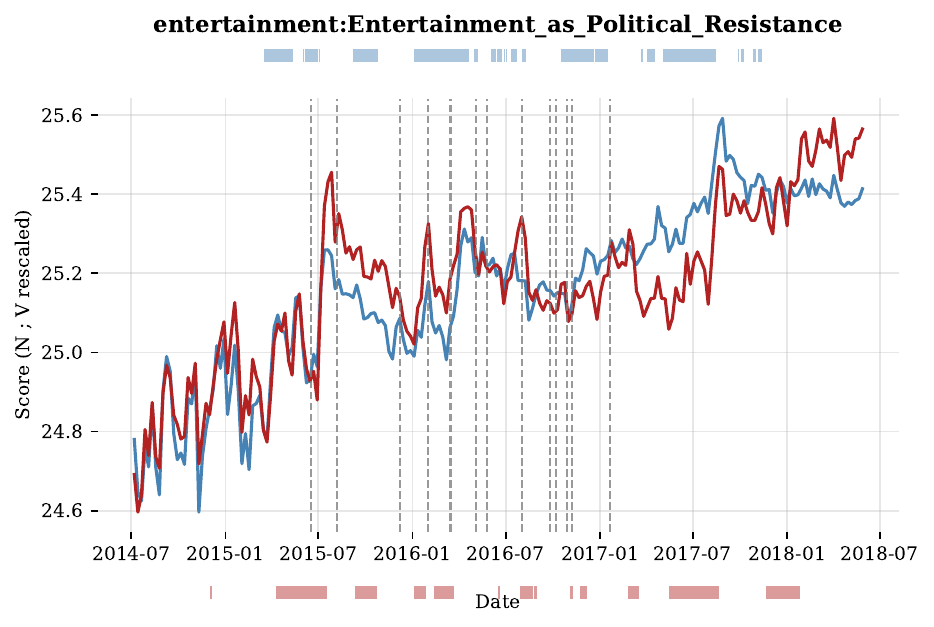}
    \caption{Narrative 2}
  \end{subfigure}

  \vspace{0.5em}
  \begin{subfigure}[t]{0.49\textwidth}
    \centering
    \includegraphics[width=\linewidth]{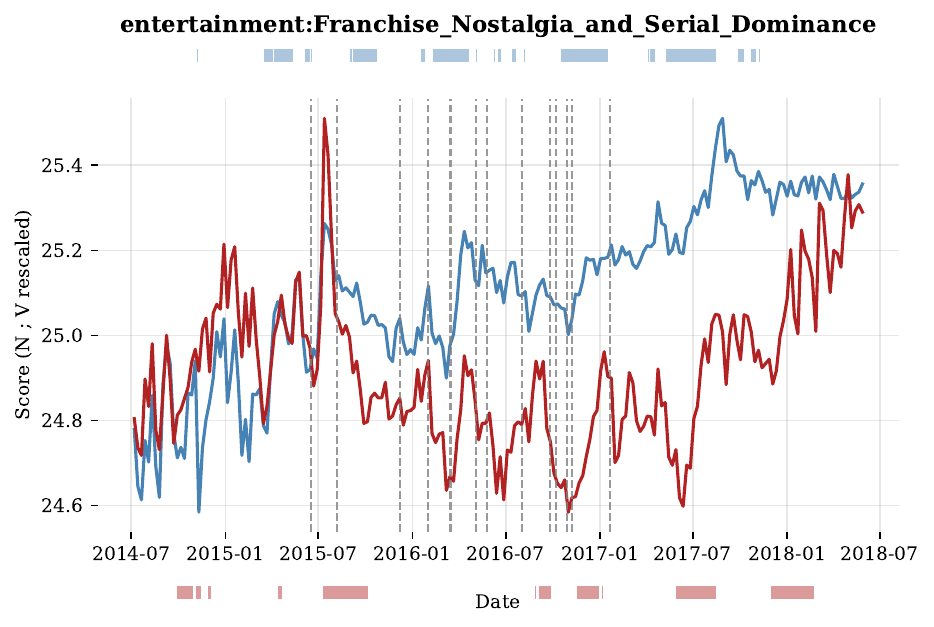}
    \caption{Narrative 3}
  \end{subfigure}
  \hfill
  \begin{subfigure}[t]{0.49\textwidth}
    \centering
    \includegraphics[width=\linewidth]{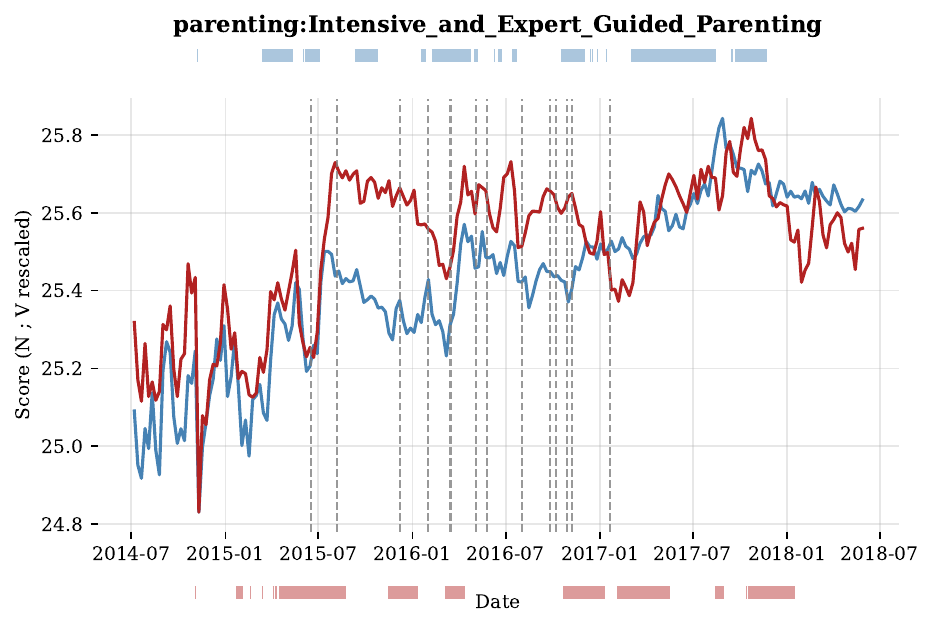}
    \caption{Narrative 4}
  \end{subfigure}

  \vspace{0.5em}
  \begin{subfigure}[t]{0.49\textwidth}
    \centering
    \includegraphics[width=\linewidth]{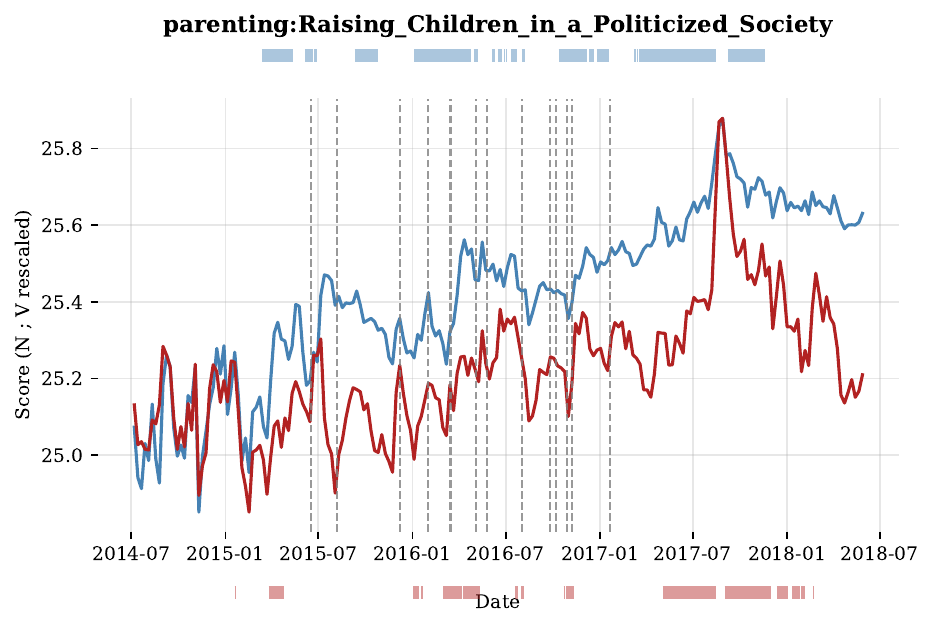}
    \caption{Narrative 5}
  \end{subfigure}
  \hfill
  \begin{subfigure}[t]{0.49\textwidth}
    \centering
    \includegraphics[width=\linewidth]{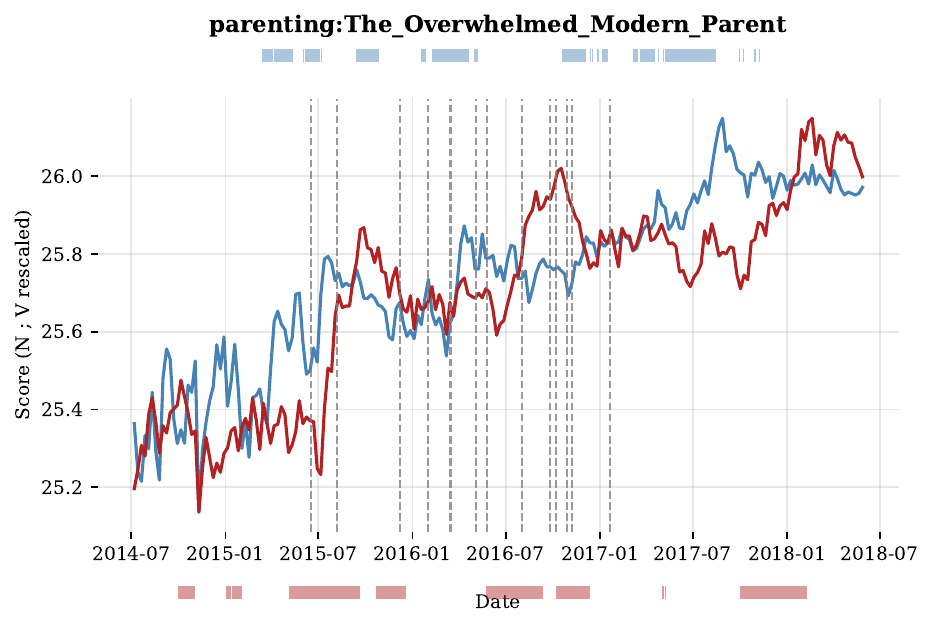}
    \caption{Narrative 6}
  \end{subfigure}

  \caption{Detected ruptures (FDR-corrected) on N (solid blue) and V rescaled
  (dashed red). The bars above and below each panel mark the rupture periods
  detected on N and V respectively. Dashed gray vertical lines indicate
  exogenous events.}
  \label{fig:ruptures}
\end{figure}

\begin{figure}[!htbp]
  \ContinuedFloat
  \centering
  \includegraphics[width=0.6\textwidth]{legend.pdf}\par\vspace{0.6em}
  \begin{subfigure}[t]{0.49\textwidth}
    \centering
    \includegraphics[width=\linewidth]{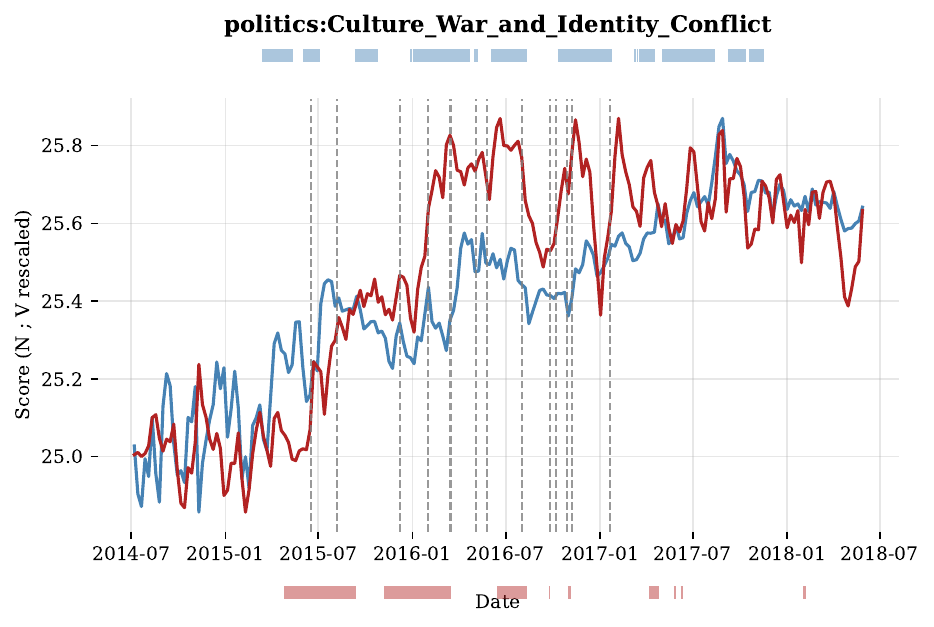}
    \caption{Narrative 7}
  \end{subfigure}
  \hfill
  \begin{subfigure}[t]{0.49\textwidth}
    \centering
    \includegraphics[width=\linewidth]{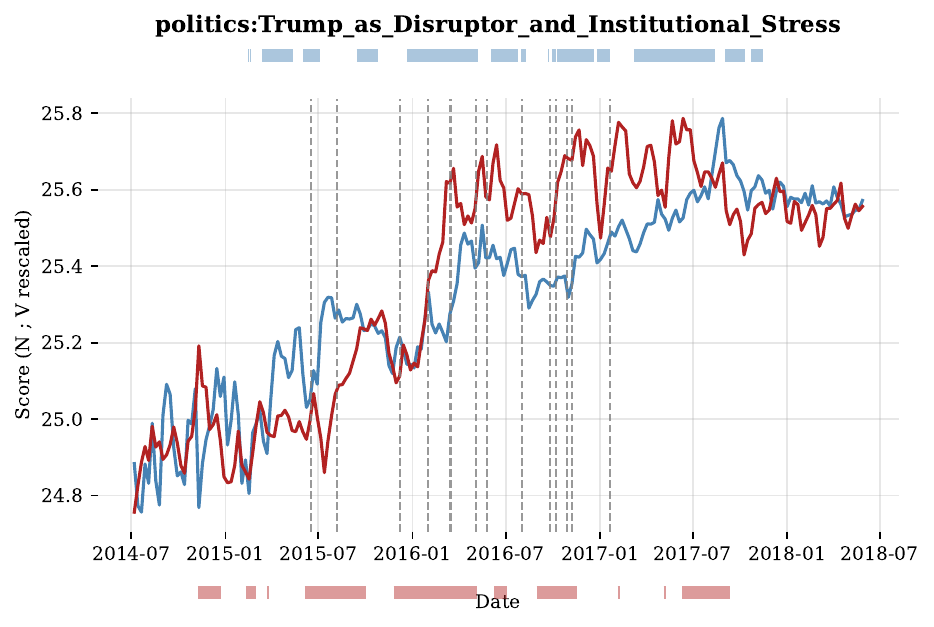}
    \caption{Narrative 8}
  \end{subfigure}

  \vspace{0.5em}
  \begin{subfigure}[t]{0.49\textwidth}
    \centering
    \includegraphics[width=\linewidth]{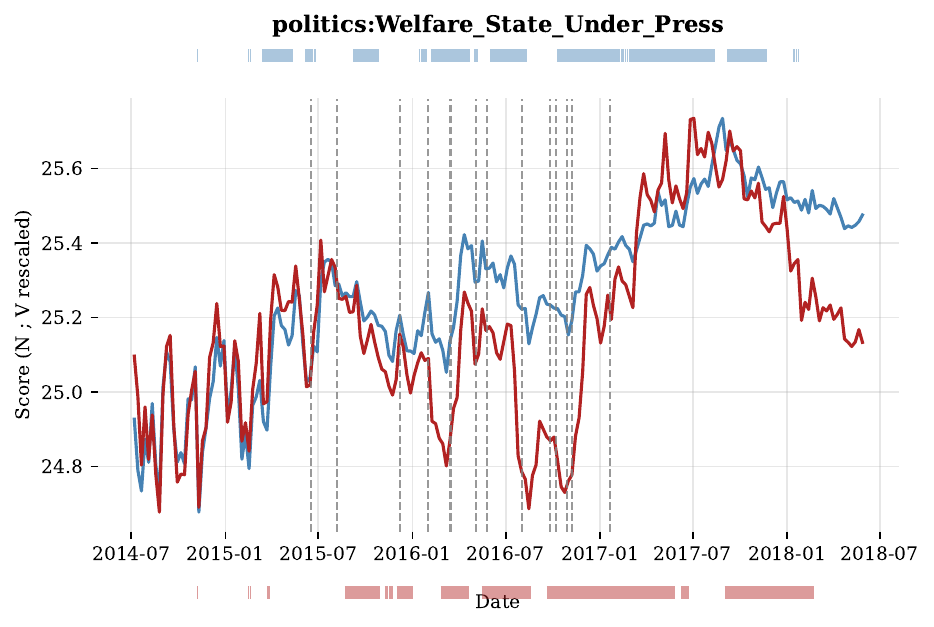}
    \caption{Narrative 9}
  \end{subfigure}
  \hfill
  \begin{subfigure}[t]{0.49\textwidth}
    \centering
    \includegraphics[width=\linewidth]{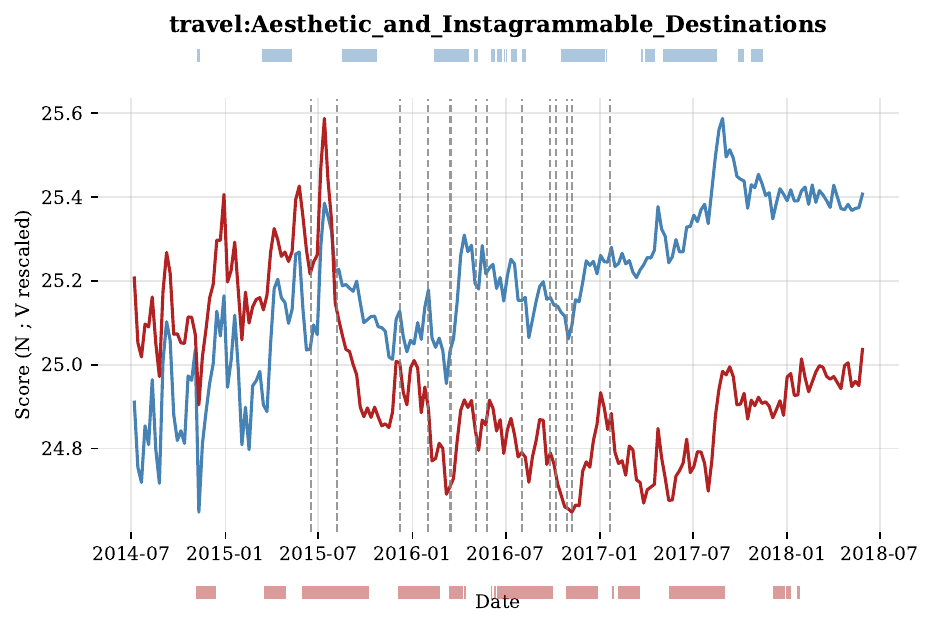}
    \caption{Narrative 10}
  \end{subfigure}

  \vspace{0.5em}
  \begin{subfigure}[t]{0.49\textwidth}
    \centering
    \includegraphics[width=\linewidth]{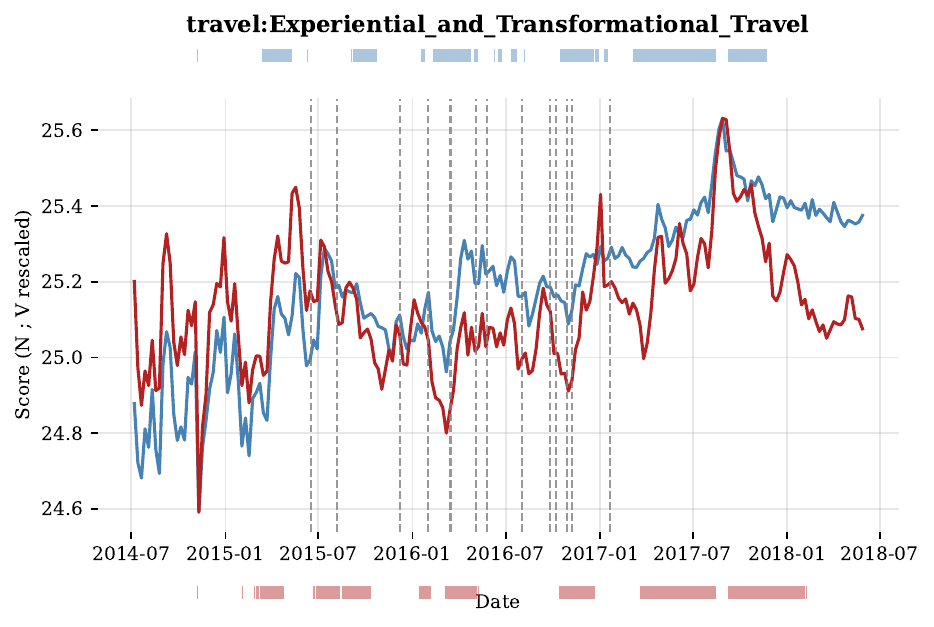}
    \caption{Narrative 11}
  \end{subfigure}
  \hfill
  \begin{subfigure}[t]{0.49\textwidth}
    \centering
    \includegraphics[width=\linewidth]{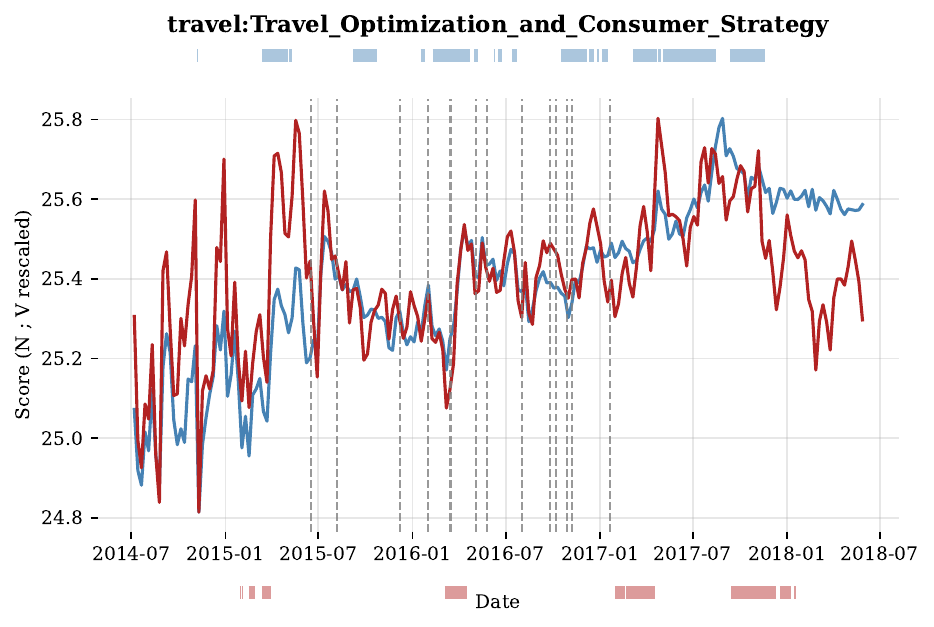}
    \caption{Narrative 12}
  \end{subfigure}

  \caption[]{Detected ruptures (continued).}
\end{figure}

\end{document}